\documentclass[twocolumn]{aastex63}

\shorttitle{Planck detection of HSC protoclusters at $ z\sim4$}
\shortauthors{Kubo et al.}
\definecolor{grey}{rgb}{0.6, 0.6, 0.6}

\def \lsim {\hspace{0.3em}\raisebox{0.4ex}{$<$}\hspace{-0.75em}\raisebox{-.7ex}{$\sim$}\hspace{0.3em}}
\def \gsim {\hspace{0.3em}\raisebox{0.4ex}{$>$}\hspace{-0.75em}\raisebox{-.7ex}{$\sim$}\hspace{0.3em}}
\usepackage{graphicx}
\begin{document}

\title{Planck far-infrared detection of Hyper Suprime-Cam protoclusters at $\bf z\sim4$: hidden AGN and star formation activity}

\email{mariko.kubo@nao.ac.jp}

\author[0000-0000-00000-0000]{Mariko Kubo}
\affil{National Astronomical Observatory of Japan 2-21-1, Osawa, Mitaka, Tokyo, 181-8588, Japan}

\author[0000-0000-00000-0000]{Jun Toshikawa}
\affil{Institute for Cosmic Ray Research, The University of Tokyo, Kashiwa, Chiba 277-8582, Japan}

\author[0000-0000-00000-0000]{Nobunari Kashikawa}
\affil{Department of Astronomy, School of Science, The University of Tokyo, 7-3-1 Hongo, Bunkyo, Tokyo 113-0033, Japan}

\author[0000-0000-00000-0000]{Yi-Kuan Chiang}
\affil{Department of Physics \& Astronomy, Johns Hopkins University, 3400 N. Charles Street, Baltimore, MD 21218, USA}

\author[0000-0000-00000-0000]{Roderik Overzier}
\affil{Observat\'orio Nacional, Rua Jos\'eCristino, 77. CEP 20921-400, Sa\~o Crist\'ova\~o, Rio de Janeiro-RJ, Brazil}
\affil{Institute of Astronomy, Geophysics and Atmospheric Sciences, University of S\~ao Paulo, S\~ao Paulo, SP 05508-090, Brazil}

\author[0000-0000-00000-0000]{Hisakazu Uchiyama}
\affil{National Astronomical Observatory of Japan
2-21-1, Osawa, Mitaka, Tokyo, 181-8588, Japan}
\affil{Department of Astronomical Science, Graduate University 
for Advanced Studies (SOKENDAI), Mitaka, Tokyo 181-8588, Japan}

\author[0000-0000-00000-0000]{David L. Clements}
\affil{Astrophysics Group, Imperial College London, Blackett Laboratory, Prince Consort Road,
London SW7 2AZ, UK}

\author[0000-0000-00000-0000]{David M. Alexander}
\affil{Centre for Extragalactic Astronomy, Department of Physics, Durham University, Durham DH1 3LE, UK}

\author[0000-0000-00000-0000]{Yuichi Matsuda}
\affil{National Astronomical Observatory of Japan 2-21-1, Osawa, Mitaka, Tokyo, 181-8588, Japan}
\affil{Department of Astronomical Science, Graduate University 
for Advanced Studies (SOKENDAI), Mitaka, Tokyo 181-8588, Japan}

\author[0000-0000-00000-0000]{Tadayuki Kodama}
\affil{Astronomical Institute, Tohoku University, Sendai, Miyagi 980-8578, Japan}

\author[0000-0000-00000-0000]{Yoshiaki Ono}
\affil{Institute for Cosmic Ray Research, The University of Tokyo, Kashiwa, Chiba 277-8582, Japan}

\author[0000-0000-00000-0000]{Tomotsugu Goto}
\affil{National Tsing Hua University No. 101, Section 2, Kuang-Fu Road, Hsinchu, Taiwan 30013}

\author[0000-0000-00000-0000]{Tai-An Cheng}
\affil{Astrophysics Group, Imperial College London, Blackett Laboratory, Prince Consort Road,
London SW7 2AZ, UK}

\author[0000-0000-00000-0000]{Kei Ito}
\affil{Department of Astronomical Science, Graduate University 
for Advanced Studies (SOKENDAI), Mitaka, Tokyo 181-8588, Japan}

%% Note that the \and command from previous versions of AASTeX is now
%% depreciated in this version as it is no longer necessary. AASTeX 
%% automatically takes care of all commas and "and"s between authors names.

%% AASTeX 6.1 has the new \collaboration and \nocollaboration commands to
%% provide the collaboration status of a group of authors. These commands 
%% can be used either before or after the list of corresponding authors. The
%% argument for \collaboration is the collaboration identifier. Authors are
%% encouraged to surround collaboration identifiers with ()s. The 
%% \nocollaboration command takes no argument and exists to indicate that
%% the nearby authors are not part of surrounding collaborations.

%% Mark off the abstract in the ``abstract'' environment. 
\begin{abstract}
We perform a stacking analysis of   
{\it Planck}, {\it AKARI}, Infrared Astronomical Satellite ($IRAS$), 
Wide-field Infrared Survey Eplorer ($WISE$), and {\it Herschel} 
images of the largest number of (candidate) protoclusters at $z\sim3.8$
selected from the Hyper Suprime-Cam Subaru Strategic Program (HSC-SSP). 
Stacking the images of the $179$ candidate protoclusters, 
the combined infrared (IR) emission of the protocluster galaxies 
in the observed $12-850~\mu$m wavelength range is successfully detected with $>5\sigma$ significance (at $Planck$). 
This is the first time that the average IR
spectral energy distribution (SED) of a protocluster has been constrained at $z\sim4$.
The observed IR SEDs of the protoclusters exhibit significant excess emission in the mid-IR 
compared to that expected from typical star-forming galaxies (SFGs). 
They are reproduced well using SED models of 
intense starburst galaxies with warm/hot dust heated by young stars, 
or by a population of active galactic nuclei (AGN)/SFG composites. 
For the pure star-forming model, a total IR (from 8 to 1000 $\mu$m) 
luminosity of $19.3_{-4.2}^{+0.6}\times10^{13}~L_{\odot}$ and 
a star formation rate (SFR) of $16.3_{-7.8}^{+1.0}\times10^3~M_{\odot}$ yr$^{-1}$ are found
whereas for the AGN/SFG composite model, 
$5.1_{-2.5}^{+2.5}\times10^{13}~L_{\odot}$ and $2.1^{+6.3}_{-1.7}\times10^3~M_{\odot}$ yr$^{-1}$ are found.
Uncertainty remaining in the total SFRs; however, 
the IR luminosities of the most massive protoclusters are likely to continue increasing up to $z\sim4$. 
Meanwhile, no significant IR flux excess is observed around optically selected QSOs 
at similar redshifts, which confirms previous results. 
Our results suggest that the $z\sim4$ protoclusters trace 
dense, intensely star-forming environments that may also host 
obscured AGNs missed by the selection in the optical.  

\end{abstract}

%% Keywords should appear after the \end{abstract} command. 
%% See the online documentation for the full list of available subject
%% keywords and the rules for their use.

%% From the front matter, we move on to the body of the paper.
%% Sections are demarcated by \section and \subsection, respectively.
%% Observe the use of the LaTeX \label
%% command after the \subsection to give a symbolic KEY to the
%% subsection for cross-referencing in a \ref command.
%% You can use LaTeX's \ref and \label commands to keep track of
%% cross-references to sections, equations, tables, and figures.
%% That way, if you change the order of any elements, LaTeX will
%% automatically renumber them.

%% We recommend that authors also use the natbib \citep
%% and \citet commands to identify citations.  The citations are
%% tied to the reference list via symbolic KEYs. The KEY corresponds
%% to the KEY in the \bibitem in the reference list below. 

%----------------------------------------------
\section{Introduction} \label{sec:intro}

Overdense regions at a high redshift known as protoclusters 
are plausible progenitors of clusters of galaxies today
and thus important targets to prove the formation history of galaxy clusters 
and the giant ellipticals/brightest cluster galaxies (BCGs) therein. 
Today, many protoclusters are found using the surveys of overdensities of 
Lyman break galaxies (LBGs; e.g., \citealt{1998ApJ...492..428S,2012ApJ...750..137T,2018ApJ...856..109O}), 
Ly$\alpha$ emitters (LAEs; e.g., \citealt{2000A&A...358L...1K,2000ApJ...532..170S,2004AJ....128.2073H,2007A&A...461..823V,2019PASJ...71L...2K, 2019ApJ...879...28H,2019arXiv190209555H}), 
H$\alpha$ emitters (HAEs; e.g., \citealt{2004A&A...428..793K,2011PASJ...63S.415T,2011MNRAS.416.2041M}),
and color selection with $Spitzer$ infrared array camera (IRAC) 
\citep{2012ApJ...749..169G,2013ApJ...769...79W,2014ApJ...786...17W,2016ApJ...830...90N,2018ApJ...859...38N}. 

The protoclusters at $z=2-4$, the peak of the cosmic star formation
density history \citep{2014ARA&A..52..415M}, are paticularly important targets
to constrain the star formation history of cluster galaxies. 
A bunch of massive galaxies has already appeared
in the protoclusters at $z\lsim3$ reported by
detecting protocluster galaxies at near-infrared (NIR) such as H$\alpha$ emitters 
(e.g., \citealt{2011MNRAS.416.2041M,2013MNRAS.428.1551K}), 
color selected massive galaxies (e.g., \citealt{2007MNRAS.377.1717K,2012ApJ...750..116U,2013ApJ...778..170K,2016ApJ...830...90N,2018ApJ...859...38N}), 
including passively evolving galaxies \citep{2013ApJ...778..170K,2016ApJ...830...90N,2018ApJ...859...38N,2018MNRAS.481.5630S, 2019ApJ...871...83S}. 
During the last decade, overdensities of dusty star-forming galaxies (DSFGs), 
which have a star formation rate (SFR) of $\sim$ several $100~M_{\odot}$ yr$^{-1}$  or more
but whose ultraviolet (UV) light is absorbed by dust and re-emitted 
as thermal emission in the infrared (IR) (e.g.,\citealt{2014PhR...541...45C}), 
in protoclusters have been found via deep observations at a mid-IR to mm wavelength
using $Spitzer$, $Herschel$ and ground-based sub-milimeter telescopes/arrays
(e.g., \citealt{2009Natur.459...61T, 2016MNRAS.460.3861K, 2016ApJ...828...56W,2017ApJ...835...98U,2018PASJ...70...65U, 2018Natur.556..469M,2018ApJ...856...72O,2018A&A...620A.202A, 2019ApJ...872..117G,2019MNRAS.486.4304S,2019arXiv190209555H}). 
Such DSFGs are likely main progenitors of cluster giant ellipticals 
because they are compatible with the instantaneous star formation history 
expected for them. 
Particularly at the cores of protoclusters, a substantial number of the star formation activities
are hidden in the optical but detectable in the IR
(e.g., \citealt{2016ApJ...828...56W,2018PASJ...70...65U, 2018Natur.556..469M,2018ApJ...856...72O}). 
The deep $X$-ray observations show AGN overdensities and an enhanced AGN fraction in protoclusters 
that indicate the environmental dependence of super massive black hole (SMBH) growth 
though a wide range of values has been reported 
\citep{2009ApJ...691..687L, 2010MNRAS.407..846D, 2013ApJ...765...87L, 2013ApJ...778..170K,2017MNRAS.470.2170K, 2019ApJ...874...54M}. 

Although there is an increasing number 
of studies of protoclusters, it remains difficult to obtain robust statistics
because they are quite rare ($\sim1.5$ deg$^{-2}$ at $z\sim4$ in \citealt{2016ApJ...826..114T}). 
Previous studies have concentrated on high redshift radio galaxies (HzRGs) or luminous QSOs
(e.g., \citealt{2007A&A...461..823V,2011MNRAS.416.2041M,2019PASJ...71L...2K}), 
which are thought to have evolved into the BCGs of today in general 
while some protoclusters have been found by chance (e.g., \citealt{1998ApJ...492..428S,2012ApJ...748L..21S,2012ApJ...750..137T,2016ApJ...822....5I}).
The selection bias on them is not clear. 
In addition, the properties of the known protoclusters vary widely
even when they are at the same redshift. 
Thus there is a strong need for a large statistical study of protoclusters 
to investigate their typical properties and variations 
as a function of cluster mass and redshift. 

The on-going wide and deep optical imaging survey 
by the Hyper Suprime-Cam Subaru Strategic Program 
(HSC-SSP; \citealt{2018PASJ...70S...4A}) is performing a wide and uniform survey 
of protoclusters at $z=2-7$ \citep{2018PASJ...70S..12T}. 
At this point, $\approx180$ protocluster candidates at $z\sim3.8$ 
have been found based on the overdensities of LBGs.
According to past protocluster  studies, 
the properties of protoclusters observed in the optical
are expected to be only ``the tip of the iceberg". 
However, it is difficult to conduct multi-band follow up 
observations for such a large catalog of candidates. 
Particularly, spectral energy distributions (SEDs) in the IR
are necessary to constrain the SFR and AGN activities obscured by dust.
The Atacama Large Millimeter Array (ALMA) 
can cover only a portion of the entire IR SEDs.
Unfortunately, currently space telescopes capable 
of detecting the redshifted mid-far IR emission of galaxies at high redshift are not available. 

Here we perform the statistical study of the IR properties of the protoclusters 
by using the archival IR all sky maps. 
Lately, plausible clusters of DSFGs at $z=2$ to $4$
detected as point sources on {\it Planck} high frequency instrument
(HFI) sky images has been reported 
(e.g., \citealt{2014MNRAS.439.1193C,2016MNRAS.461.1719C, 2018MNRAS.476.3336G}). 
The spatial resolution and detection limit of the $Planck$ HFI images
are too low to discretely identify galaxies at high redshift;
however they are useful to evaluate the average total sum flux of protocluster galaxies, 
though it is difficult to individually detect protoclusters at $z\sim4$. 
In this study, the average of all the IR fluxes 
from a protocluster at $z\sim3.8$ is shown for the first time, 
by stacking the publicly available archival IR images taken by 
{\it Planck} \citep{2011A&A...536A...1P}, {\it AKARI} \citep{2007PASJ...59S.369M}, 
Infrared Astronomical Satellite ({\it IRAS:} \citealt{1984ApJ...278L...1N}),
Wide-Field Infrared Survey Explorer ({\it WISE:} \citealt{2010AJ....140.1868W}),
and the Herschel Astrophysical Terahertz Large Area Survey 
($H$-ATLAS data release 1, \citealt{2016MNRAS.462.3146V})
of the largest catalog of the candidate protoclusters 
at $z\sim3.8$ selected from the HSC-SSP survey. 

This paper is organized as follows: 
in Section 2, the HSC-SSP protocluster catalog and archival IR data are described, 
in Section 3, the stacking analysis methods are described, 
in Section 4, the results are presented, 
and in Section 5, the findings are discussed. 
Throughout the paper, a $\Lambda$CDM cosmology is adopted 
with $H_0=70$ km s$^{-1}$ Mpc$^{-1}$, 
$\Omega_{\Lambda} = 0.7$ and $\Omega_m = 0.3$.

%% Note that the \setcounter and \renewcommand are needed here because
%% this example is using a mix of deluxetable and tabular.  Here the
%% deluxetable counters are set with \tablenum but the situation is a bit
%% more complex for tabular.  Use the first command to set the Table number
%% to ONE LESS than it should be.  The next command will auto increment it
%% to the desired number.
\begin{figure*}
\centering
\includegraphics[width=160mm]{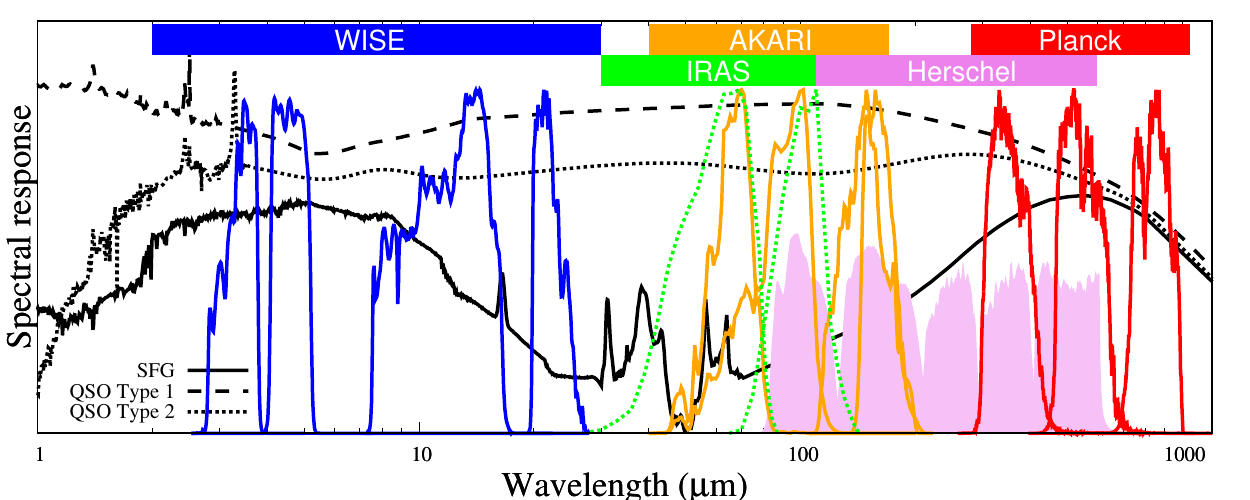}
\caption{
Filter transmission curves for the IR sky surveys used in this study.
The red curves at $350-850~\mu$m show {\it Planck} 353, 545, and 857 GHz.
The orange curves at $40-200~\mu$m show 
{\it AKARI} {\it N60}, {\it WIDE-S}, {\it WIDE-L}, and {\it N160}. 
The green dotted curves show {\it IRAS} 60 and 100 $\mu$m.
The violet filled curves show {\it Herschel} 100, 160, 250, 350, and 500 $\mu$m. 
The blue curves at $3-30~\mu$m show {\it WISE} {\it W1}, {\it W2}, {\it W3}, and {\it W4}. 
The black thick solid, dashed and dotted curves show the SEDs of a SFG, Type-1 QSO and  Type-2 QSO 
from the SED library by \citet{2006ApJ...642..673P} shifted to $z=3.8$, respectively.
}
\label{fig:bandpass}
\end{figure*}

\section{Data}
\label{sec:data_sample}
\subsection{Protocluster catalog}

We use the protocluster catalog at $z\sim3.8$ obtained via 
a systematic search for high-z protoclusters based on the HSC-SSP survey
\citep{2018PASJ...70S...4A} in \citet{2018PASJ...70S..12T}.
HSC is the prime focus camera with a large field-of-view 
(FoV) of 1.8 deg$^2$ and high sensitivity 
on Subaru Telescope \citep{2012SPIE.8446E..0ZM}. 
The HSC-SSP survey is an on-going multi-color ($griz$ + narrow band) 
survey consisting of three layers; 
the ultradeep (UD; 3.5 deg$^2$, $i\sim28$ mag), 
deep (26 deg$^2$, $i\sim27$ mag) 
and wide (1400 deg$^2$, $i\sim26$ mag) layers.
\citet{2018PASJ...70S..12T} constructed a catalog of LBGs 
based on the $gri$-band images (hereafter the $g$-dropout galaxies) 
over an area of 121 deg$^2$ 
of the HSC-SSP wide survey. 
Their color cut is sensitive to galaxies in the redshift range of $3.3\lsim z\lsim 4.2$. 
Then, they measured the surface density of the $g$-dropout galaxies selected 
down to a limiting magnitude of $i=25.0$ mag
within an aperture of $1.8$ arcmin (0.75 Mpc physical).
They then, selected the regions with an overdensity significance of $>4\sigma$ 
as protocluster candidates. 
This radius corresponds to the typical extent of the regions 
that will collapse into a single massive halo with a halo mass $M_h>10^{14}~M_{\odot}$ 
by $z=0$ predicted in the cosmological numerical simulations (e.g., \citealt{2013ApJ...779..127C}).
Because the redshift range of the $g$-dropout galaxies is large, 
some of the protocluster candidates are probably spurious because of projection effects. 
They quantified this possible contamination using simulations, 
finding that at a $4\sigma$ threshold,  
approximately $>76$~\% of the candidate protoclusters are expected to evolve into 
massive galaxy clusters.
According to the correlation function analysis in \citet{2018PASJ...70S..12T}, 
the majority of the candidates are expected to evolve into clusters of galaxies
with $M_h\geq 5\times10^{14}~M_{\odot}$, i.e., the richest clusters today. 

They finally selected 216 overdense regions. 
Of these, 37 are within $8$ arcmin from another overdense region.
Because the typical spatial extents of protoclusters drop at $\sim8$ arcmin (see Fig. 8 of \citealt{2016ApJ...826..114T}), 
they can be substructures of larger overdense regions.
After merging these regions, they identified 179 unique protocluster candidates. 
In the following stacking analysis, 
the archival IR images centered on these 216 density peaks
of the $g$-dropout galaxies in each protocluster candidate are cut out. 

\subsection{IR all sky surveys}
We perform a stacking analysis of the protoclusters 
using the publicly available archival IR images. 
We use {\it Planck}, {\it AKARI}, {\it IRAS}, and {\it WISE} all sky survey,  
and $H$-ATLAS.  
Fig. \ref{fig:bandpass} shows their filter transmission curves. 
They cover a large portion of the IR SEDs of galaxies at $z\sim4$. 
Table \ref{tab:data} in the Appendix summarizes the central wavelengths, 
full width at half-maximum (FWHM) of the point spread functions 
(PSFs), point source detection limits,
and expected sky noises on the stacked images. 
In the following, we provide a brief description of the archival IR images used here.

\subsubsection{Planck} 
We use the 353, 545, and 857 GHz images taken 
by the {\it Planck} HFI 
in the $Planck$ legacy archive\footnote{https://www.cosmos.esa.int/web/planck}. 
They cover $350$ to $850~\mu$m 
with a FWHM of the PSF from $4.2$ to $4.9$ arcmin. 
Various objects are detectable on the sky images taken by {\it Planck}, 
e.g., synchrotron emission from radio sources, 
the SZ effect from galaxy clusters and Galactic dust emission.
At $z\sim3.8$, warm to cold dust emission
originating in the SFGs and AGNs shifts in 350 to 850 $\mu$m. 

The major contaminant from nearby objects is the dust emission from our Galaxy.
To reduce the contamination from Galactic dust emission, 
we use the cosmic infrared background (CIB) products \citep{2016A&A...596A.109P}\footnote{https://wiki.cosmos.esa.int/planckpla2015/index.php/CMB\\\_and\_astrophysical\_component\_maps}
in which Galactic thermal dust emission 
is subtracted using the Generalized Needlet Internal Linear 
Combination (GNILC) component separation method. 
The area heavily affected by Galactic dust emission are masked in the CIB products. 
None of the protoclusters in our catalog are in the masked area. 

\subsubsection{IRAS}
The IRAS mission 
is an all sky survey at $12-100~\mu$m \citep{1984ApJ...278L...1N}. 
Here the 60 and 100 $\mu$m images
available from NASA/IPAC Infrared science archive\footnote{http://irsa.ipac.caltech.edu} are used.
The FWHM of the PSF at 60 and 100 $\mu$m is $3.6$ and $4.2$ arcmin, respectively. 

\subsubsection{AKARI}
{\it AKARI} is the Japanese infrared astronomical satellite that 
performed all sky mapping at $9-160~\mu$m \citep{2007PASJ...59S.369M}. 
The {\it N60}, {\it WIDE-S}, {\it WIDE-L}, and {\it N160}-band images taken 
with Far InfraRed Surveyor (FIS; \citealt{2007PASJ...59S.389K}) 
available from the {\it AKARI} all-sky survey map in the public archive
\citep{2015PASJ...67...50D,2015PASJ...67...51T} are used here.
The FWHMs of the PSF on them are $1\sim1.5$ arcmin. 

\subsubsection{WISE}

The {\it WISE} \citep{2010AJ....140.1868W} 
all-sky survey mapped the sky in 3.4 ({\it W1}), 
4.6 ({\it W2}), 12 ({\it W3}), and 22 $\mu$m  ({\it W4}).  
WISE Atlas images with a FWHM of the PSF $8\sim16.5$ arcsec ($6-12$ arcsec for single exposure)
in the public archive\footnote{http://wise2.ipac.caltech.edu/docs/release/allsky/} are used. 
On the {\it WISE} images, 
not only Galactic dust emission but also many foreground stars/galaxies 
are the major cause of the noise for our stacking analysis. 

\subsection{$H$-ATLAS}

The $H$-ATLAS surveyed 161 deg$^2$ of the Galaxy Mass and Assembly (GAMA) field 
at 100, 160, 250, 350, and 500 $\mu$m. 
The 93 protoclusters in our catalog are enclosed in $H$-ATLAS. 
Here RAW images at 250, 350, and 500 $\mu$m 
provided in the public archive\footnote{http://www.h-atlas.org} are used. 
$H$-ATLAS also provides background subtracted 
(BACKSUB) images at 100, 160, 250, 350, and 500 $\mu$m;
however, they are not used because of the over sky subtraction problem described in Section 3.4. 
Because only BACKSUB images are available for 100 and 160 $\mu$m, 
they are excluded from the stacking analysis of the protoclusters. 
The signal to noise (S/N) ratios for 4-arcmin diameter photometries
on stacked images of $H$-ATLAS are lower than those of the $Planck$ images 
because of the two-fold smaller sample size and large aperture size. 
Because {\it Herschel} images have a good spatial resolution ($12-35$ arcsec), 
they are also useful to evaluate the total fluxes (Section 3.4) 
and the fluxes of the $g$-dropout galaxies and QSOs (Section 3.6). 

\begin{figure}\centering
\includegraphics[width=80mm]{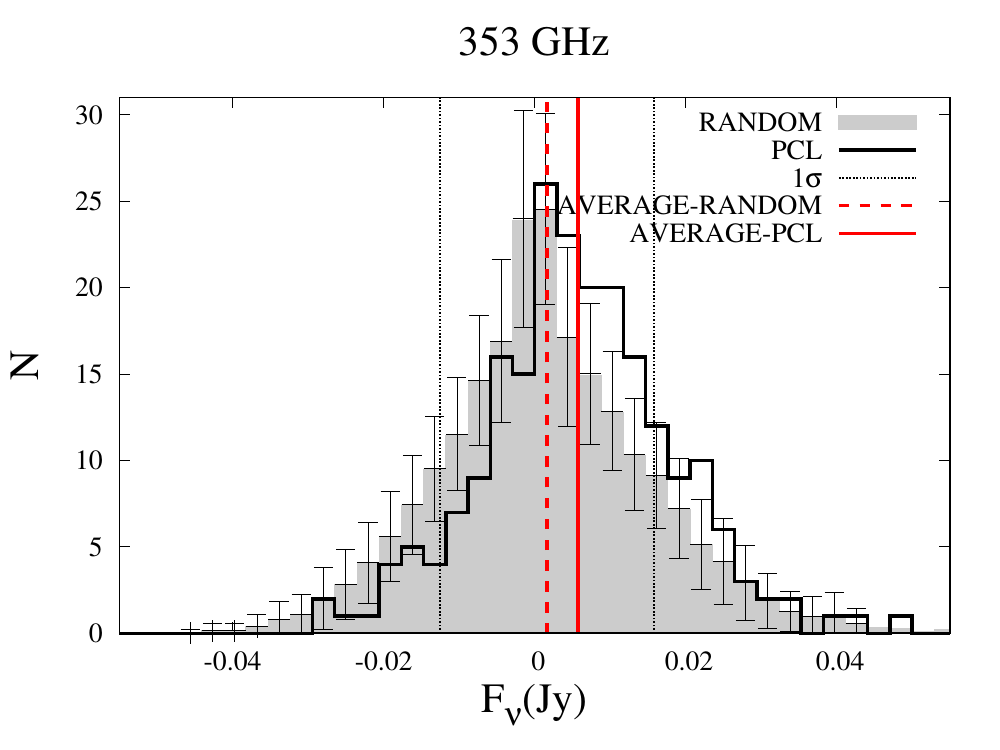}
\includegraphics[width=80mm]{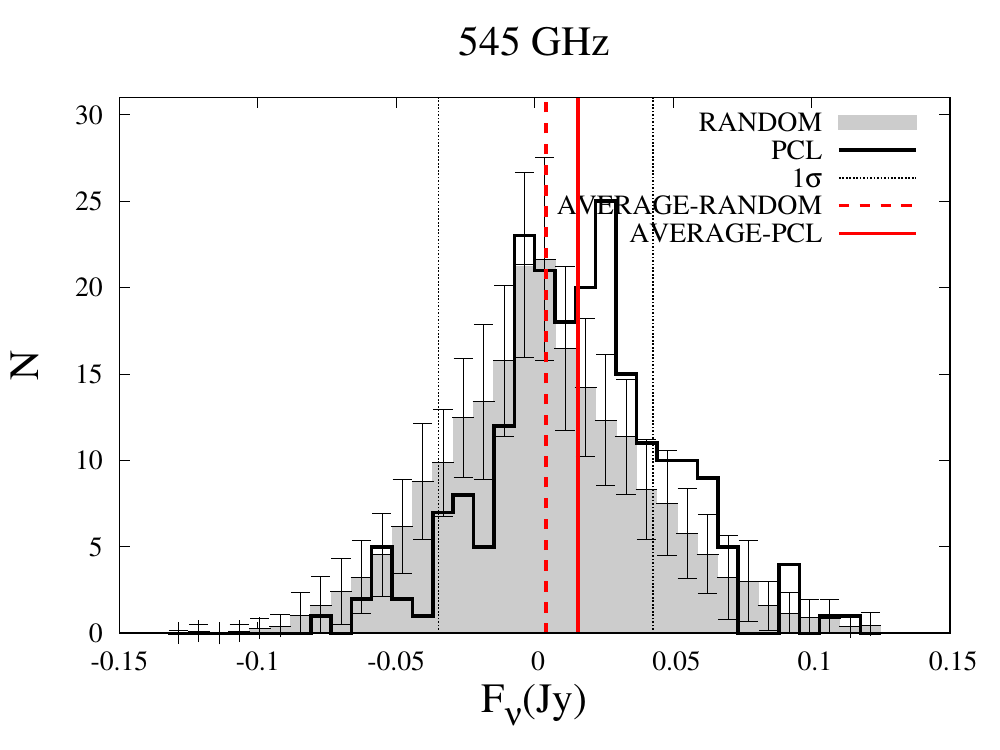}
\includegraphics[width=80mm]{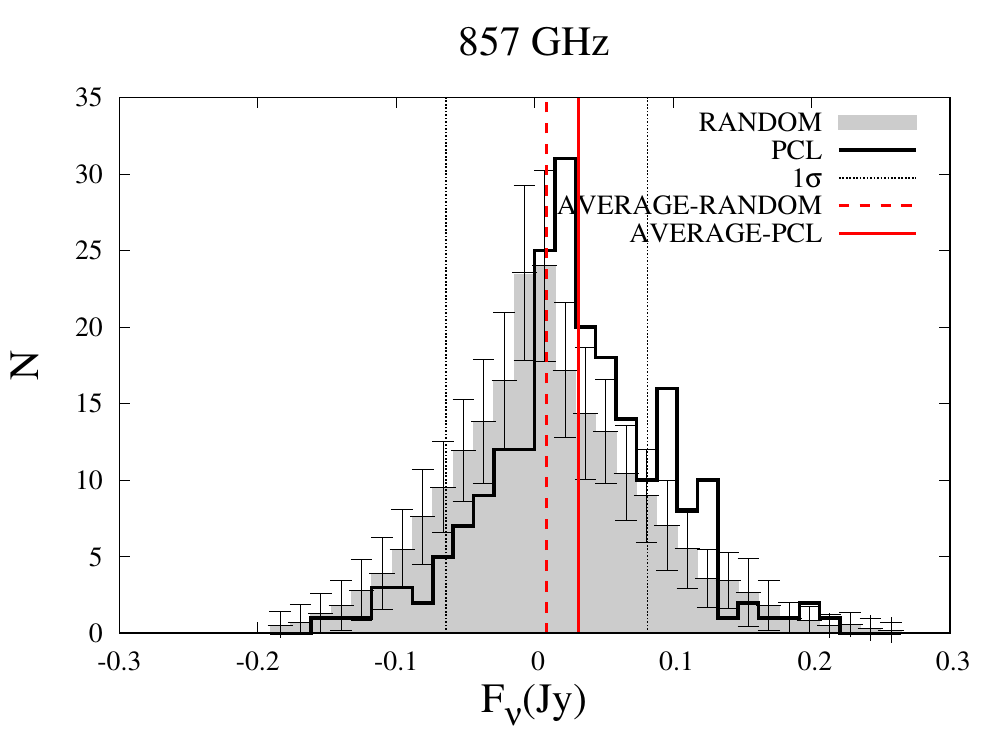}
\caption{{\it Top:} Black solid line shows the distribution 
of the flux densities at 353 GHz measured 
at the protoclusters at $z\sim3.8$. 
The gray histogram and its error show
the average and standard deviation of the flux distribution of the random sky positions. 
The vertical dotted lines show the $1\sigma$ rms noise at 353 GHz. 
The red vertical solid and dashed lines show 
the average value of the protoclusters and random sky positions, respectively.
The {\it Middle} and {\it bottom} are similar to the {\it top} panel 
but at 545 and 857 GHz, respectively. }
\label{fig:fluxdist} \end{figure}
\begin{figure}
\centering
\includegraphics[width=80mm]{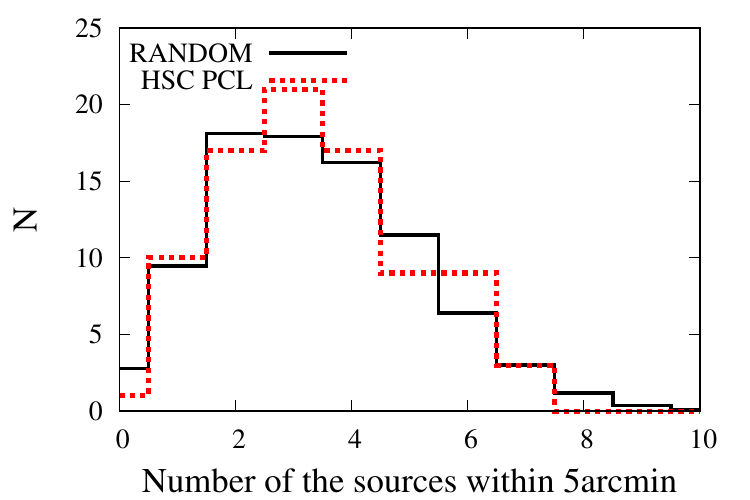}
\includegraphics[width=80mm]{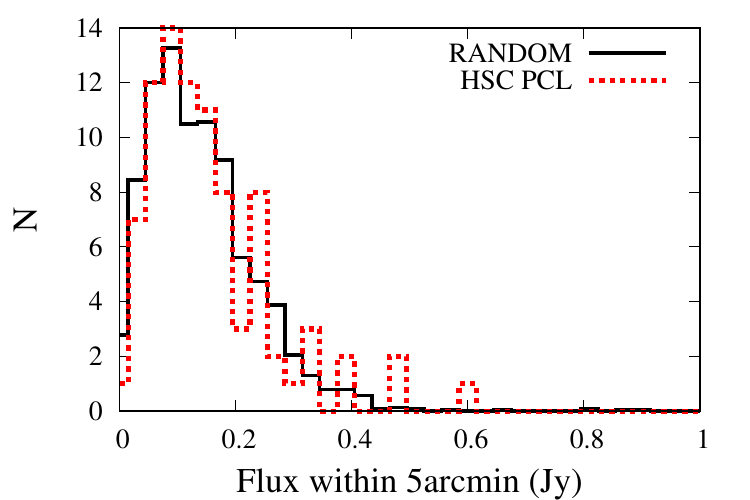}
\caption{{\it Top}: The distribution of the number count
of the $4 \sigma$ sources at 250 $\mu$m within 
the 5-arcmin diameter aperture of the protoclusters (red dashed) and random sky positions (black solid). 
{\it Bottom}: Similar to the {\it top} panel but the $x$-axis shows the sum of the fluxes of the sources within 5-arcmin diameter.}
\label{fig:herschelcount}
\end{figure}

\section{Method}
\label{sec:data_sample}

\subsection{Discrete detections of the protoclusters using {\it Planck}}

First, whether protoclusters 
are individually detected on {\it Planck} images is assessed
because e.g., \citet{2018MNRAS.476.3336G} suggests that the most actively star-bursting protoclusters 
can be detected on $Planck$. 
At the least, none of the protoclusters matches the 
the second {\it Planck} compact source catalog \citep{2016AA...596A.100P}, 
a secure catalog with high detection significance and a S/N $\gsim5$. 
To investigate the presence of fainter sources, 
the distribution of 5-arcmin diameter aperture 
flux values measured at the protoclusters and random sky positions shown in Fig. \ref{fig:fluxdist} are compared. 
The flux distribution for random sky positions 
shows the average and $1 \sigma$ standard deviation 
of the flux distributions of 216 random sky positions measured by a thousand times iteration. 
The vertical dotted lines show $\pm1\sigma$ errors.
The average fluxes of the protoclusters 
and random sky positions are indicated with red thick solid and dashed lines, respectively. 
The flux distributions of the protoclusters and random points 
are compared via the Kolmogorov-Smirnov (KS)-test,  
and $p$-values $=0.006$, 0.049, and 0.778 
are obtained at 353, 545, and 857 GHz, respectively. 
Thus, the flux distributions at 353 and 545 GHz are significantly offset from the random points. 
In addition, the centers of the flux distributions of the protoclusters 
shift brighter than those of random sky positions. 
The average fluxes at 353, 545, and 857 GHz are 6, 16, and 32 mJy, respectively,  
for the protoclusters and 2, 4, and 9 mJy for random sky positions. 
These excesses follow well the fluxes of the protoclusters measured using the stacking analysis. 

\subsection{Notes for the possibility to detect individual galaxies in the protoclusters}

$H$-ATLAS and {\it WISE} are sufficiently sensitive  
to detect the most luminous objects at $z\sim4$. 
$H$-ATLAS can detect 
a very bright SMG or a dense group of bright SMGs as a point source 
at $z\sim4$ (e.g., \citealt{2018Natur.556..469M}). 
\citet{2018ApJ...854..157F} and \citet{2018ApJ...857...31T}
reported extremely IR luminous dust obscured galaxies (DOGs) 
at $z\sim4$ detectable using {\it WISE}.  
Although such sources are quite rare, 
the possibility cannot be ignored that the protoclusters contain 
such sources detectable using {\it Herschel} and/or {\it WISE}. 

The number count and total fluxes of the sources 
detected at the protoclusters and random sky positions are compared 
using the $H$-ATLAS source catalog \citep{2016MNRAS.462.3146V} in Fig. \ref{fig:herschelcount}. 
The {\it top} panel shows the distributions 
of the number count of the objects detected above $4 \sigma$ in 250 $\mu$m 
within a 5-arcmin diameter of the protoclusters and random sky positions. 
The {\it bottom} panel is similar to the {\it top} panel except for the sum of their fluxes. 
The number and flux distributions of the protoclusters and random points 
are compared via KS-test 
and $p$-values $=0.86$ and 0.84 are found, respectively. 
Thus, there is no significant difference between the protoclusters and random sky positions. 
Similarly, there is no clear difference at 350 $\mu$m and 500 $\mu$m, 
and also {\it WISE} $W3$ and $W4$. 
However, there remains still a possibility that some extremely luminous 
protocluster members can be found using more detailed SED analysis 
of the sources detected on the $H$-ATLAS and/or {\it WISE} images.  
This is beyond the scope of this work and will be the subject of a future paper. 

\subsection{Stacking analysis using WISE, IRAS, AKARI and Planck images}

Next, we perform a stacking analysis of the protoclusters. 
Before performing the stacking analysis, we check contamination of bright foreground sources, 
subtract sky and smooth images. 
First, possible contaminants are assessed. 
Foreground objects can be resolved and detected on $AKARI$ images. 
Using {\sf SExtractor} \citep{1996A&AS..117..393B}, 
one or two sources with $1\sim3$ Jy are detected in more than one bandpass at two fields. 
Because both of the fields have no corresponding source on the $H$-ATLAS images at 100 and 160 $\mu$m, 
the sources are not foreground objects but likely noise. 
To make sure, these fields are not used. 
The other possible bright interlopers are QSOs. 
Not only QSOs themselves but also possible protoclusters around them 
at different redshifts can contaminate the signal from our targets. 
Approximately one-half of the protoclusters in our catalog are 
within 5 arcmin of the QSOs at all redshifts selected from Sloan Digital Sky Survey (SDSS). 
The effect of these QSOs is checked by performing a stacking analysis rejecting such fields. 
Because this rejection causes no change in the results, they are not rejected. 
Given the aforementioned, there is no possible bright foreground interlopers at the sky positions of our targets. 

We perform our own sky subtraction because the background sky levels 
on the archival $AKARI$, $IRAS$ and $WISE$ 
images considerably vary among the survey area of the protoclusters 
while they are nearly uniform on the {\it Planck} CIB map and $H$-ATLAS RAW images. 
For example, in the case of $AKARI$ $WideL$,
the average sky values on the archived images 
at the protoclusters vary from $0.01-0.04$ Jy/pix. 
Thus, before stacking the images, sky subtractions 
for $AKARI$, $IRAS$ and $WISE$ images are performed.
The archived $WISE$, $AKARI$, and $IRAS$ images are provided as 1.6 $\times$ 1.6 deg,  6 $\times$ 6 deg, and 12.4 $\times$ 12.4 deg cutouts, respectively. 
We evaluate the sky values on an image after masking the bright sources.
To generate object masks, we extract sources using {\sf SExtractor} \citep{1996A&AS..117..393B}. 
The sources detected above $2\sigma$ for a square of the FWHM of the PSF size region, 
and their surrounding regions for the FWHM of the PSF size radius, are masked. 
The sky is evaluated with $\approx10$ arcmin mesh 
and the sky images are generated. 
Then, the sky image is subtracted from the original image.
The sky subtraction is visually assessed to ensure it works well. 
Though sky subtraction does not work well around very bright objects, 
sky subtraction at protoclusters largely works well.

Then, we smooth the sky subtracted images such that the FWHM of the PSF $\approx4.9$ arcmin, 
similar to that on the {\it Planck} 353-GHz image. 
We cut out the images at the protoclusters 
taking the density peaks of the $g$-dropout galaxies as the centers. 
Then, we perform average stacking 
with 3$\sigma$ clippings by using the {\sf imcombine} task of {\sf IRAF}. 
If there are no value pixels at the edge of a cutout image, these pixels are ignored. 
In addition to all the protoclusters, we also perform the stacking analysis for
the brighter-half protoclusters on the {\it Planck} 857-GHz image
(see Fig. \ref{fig:fluxdist})
and the protoclusters with the overdensity significances 
of the $g$-dropout galaxies $5\sigma$ or more. 
The number of the cutout images of the brighter-half protoclusters is $N=106$, rejecting the two poor $AKARI$ fields,
and that of the $5\sigma$ overdense protoclusters is $N=67$. 
We perform 4-arcmin diameter aperture photometries 
and later convert them to the total fluxes by aperture correction. 
The average fluxes and their errors 
are the average and standard deviations measured 
via a thousand-times bootstrap resampling. 
The standard deviation does not only originate in the sky noise but also in the variation in the protoclusters. 
Finally, we apply an aperture correction to convert them into total fluxes 
measured in the next section. 

The $1\sigma$ sky noises 
expected for the stacked images are listed in Table \ref{tab:data}. 
They are the standard deviations of the flux values in 4-arcmin 
diameter apertures measured on a thousand images generated 
by stacking the images at 214 random sky positions.
The detection limits of {\it Planck} stacks
are deeper than that of the $H$-ATLAS stacks. 
Notably, when the HSC-SSP WIDE survey is completed,
a $\sim10$ times larger catalog will be available. 
Then, these stacks will be three times deeper than the current depth in the future. 

We match the PSF sizes on images by a simple Gaussian smoothing; 
however, the PSFs on the images used here are not simply similar to a Gaussian profile. 
The beam profiles on the {\it Planck} images depend on the sky positions. 
We evaluate the average beam profile of the protoclusters in 353 GHz
based on the public data-base \citep{2014A&A...571A...6P}. 
For the averaged beam profile, 
$\approx62\%$, $79\%$, and $96\%$ of the total flux of a point source is enclosed
in 4-, 5-, and 10-arcmin diameter apertures, respectively. 
For a Gaussian profile with FWHM of the PSF size $=4.9$ arcmin, 
$\approx43\%$, $58\%$, and $96\%$ of the total flux of a point source is enclosed
in 4-, 5-, and 10-arcmin diameter apertures, respectively.
The PSFs on the $IRAS$ images are not the same as those on the $Planck$ images. 
Because a large Gaussian smoothing is applied on the {\it AKARI}, {\it WISE}, and {\it Herschel} images, 
their PSFs may not behave like {\it Planck} but a Gaussian profile. 
This can result in a slight inconsistency of the flux measured with different facilities. 
Practically, protoclusters should not behave similar to a point source. 
The average spatial extent of the protoclusters is measured 
using $H$-ATLAS in the next subsection.
The average radial profile of the protoclusters in $Planck$ compared to that of the PSF 
and several mock source distributions are presented in Appendix B.

\subsection{Stacking analysis of $H$-ATLAS images}
\label{sec:analysisapcorr}

\begin{figure}	\centering
\includegraphics[width=80mm]{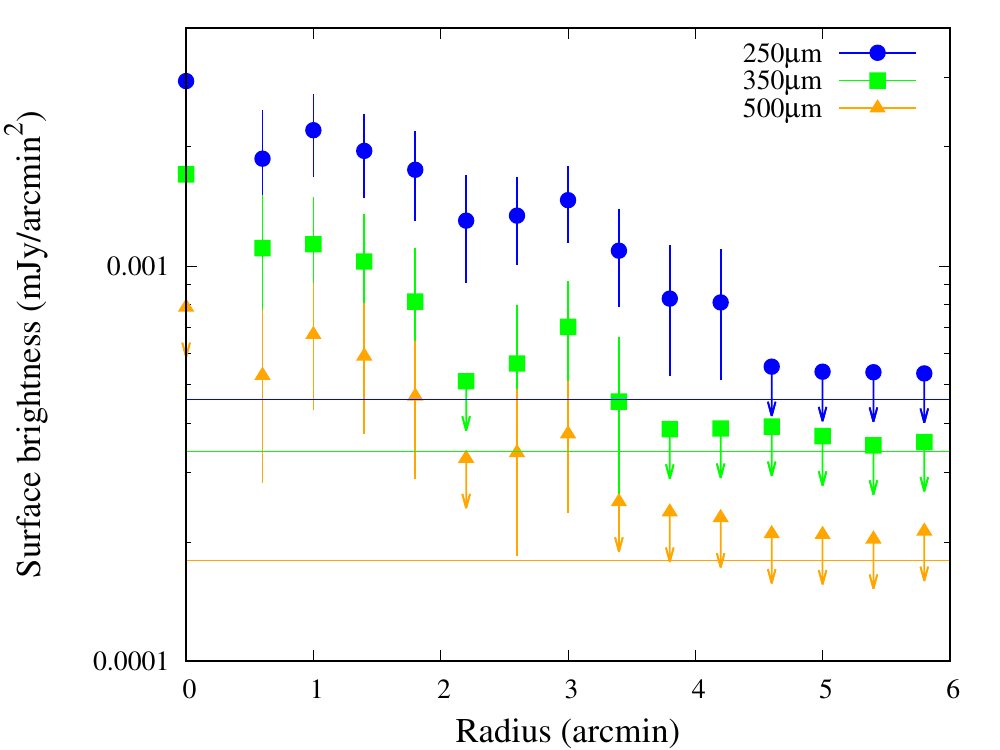}
\caption{
Average radial profile of all protoclusters measured using $Herschel$. 
The blue circles, green squares and yellow triangles show
the average radial profiles at 250, 350, and 500 $\mu$m, respectively. 
The vertical lines denote the $1\sigma$ sky noise levels measured in 1 arcmin square regions.  
Here after, a $2 \sigma$ upper limit value is shown if a signal is not detected above $2 \sigma$ significance. 
}
\label{fig:rprofile}
\end{figure}

We stack $H$-ATLAS RAW images
and those smoothed to have a FWHM of the PSF sizes similar to those of a {\it Planck} image at 353 GHz. 
With the former products, the average physical extent
and total flux of the protoclusters are limited 
while the fluxes measured on $Planck$ images 
are smoothed off because of the large PSFs and extended geometries of the protoclusters. 
Fig. \ref{fig:rprofile} shows the average radial profiles of the protoclusters 
measured at 250, 350, and 500 $\mu$m. 
Signals of the protoclusters are detected within $\approx4$ arcmin. 
For the brighter-half protoclusters, signals are detected within $\approx6$ arcmin. 
We adopt the fluxes measured at an 8-arcmin diameter (12-arcmin for the brighter-half) 
as the lower limits of the total fluxes. 

\begin{figure}	\centering
\includegraphics[width=85mm]{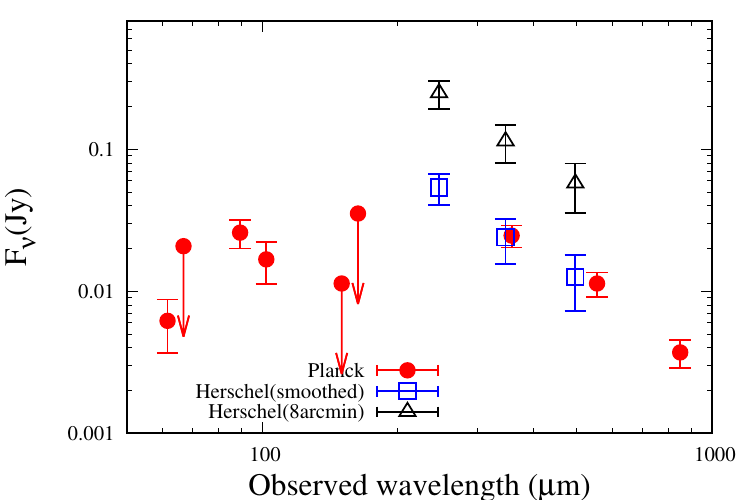}
\caption{
The red filled circles show the fluxes measured on the stacked images of {\it Planck}, {\it AKARI}, and {\it IRAS}. 
The blue open squares show the fluxes measured on the stacked images 
of the {\it Herschel} images smoothed 
to match the PSF sizes with those of {\it Planck} at 353 GHz. 
For these two, we perform 4-arcmin diameter aperture photometries. 
The black open triangles show the fluxes measured
using an 8-arcmin diameter aperture on the stacked images 
of the {\it Herschel} images without smoothing. 
}
\label{fig:hpai}
\end{figure}

The black triangles and blue squares in Fig. \ref{fig:hpai} 
show the total fluxes measured by stacking $Herschel$ images
and 4-arcmin diameter aperture fluxes measured 
by stacking $Herschel$ images matched the PSF sizes to $Planck$ 353 GHz, respectively.
The fluxes measured using the PSF-matched $Herschel$ images 
match quite well those measured using $Planck$ images. 
The average flux ratios  
are $4.7\pm2.2$ and $6.6\pm2.1$ for all and the brighter-half protoclusters, respectively. 
Here, these ratios are adopted as the aperture correction factors. 
For the $5 \sigma$ overdensity protoclusters, 
we apply the aperture correction factor for all protoclusters. 
The aperture correction factors for various source geometries expected for protoclusters
are simulated as presented in Appendix B. 
These aperture correction factors are consistent 
with those of the simulated protocluster geometries. 

We note that background subtracted products (BACKSUB) of $H$-ATLAS 
are not suitable for the stacking analysis of the protoclusters 
because the fluxes of our targets are greatly reduced as a consequence of their sky subtraction. 
For example, at 250 $\mu$m, the flux of the protoclusters measured using RAW images 
is two times greater than that using BACKSUB images. 
This result is perhaps because sky subtraction is performed at a scale smaller 
than the typical extent of the protoclusters and the subtracted 
sky values are similar to the protoclusters fluxes.  
The difference between the BACKSUB and RAW images 
is less ($<10\%$) for $g$-dropout galaxies and QSOs (Section 3.6) 
because they are point sources and $g$-dropout galaxies are much fainter than protoclusters.

\subsection{Average optical total fluxes}
\label{sec:analysisapcorr}

Because the contamination of nearby sources is large, 
we avoid a stacking analysis in optical. 
We here limit the average total fluxes of the protocluster galaxies in the $g,r,i,z \& y$-band 
by summing the Cmodel fluxes of the $g$-dropout galaxies
in the HSC-SSP catalog public data release 1 (pdr1: \citealt{2018PASJ...70S...4A}).
First, the average total fluxes of the $g$-dropout galaxies 
with a $i\leq25.0$ mag within the 8-arcmin diameter of the protoclusters 
is measured and that of 1000 random positions. 
Then, the latter is subtracted from the former. 
Note that this is a lower limit because the contribution from the galaxies with $i>25.0$ mag are ignored.

\subsection{Stacking analysis of the $g$-dropout galaxies and SDSS QSOs}
\label{sec:analysisapcorr}

To discuss whether the optically selected objects 
can explain the entire flux of the protoclusters,  
we perform stacking analyses
of $g$-dropout galaxies and SDSS QSOs. 
Approximately $\approx238,500$ ($\approx94,000$ for $H$-ATLAS) 
$g$-dropout galaxies with an $i\leq25.0$ are used  but not within 10 arcmin of the protoclusters 
from the HSC-SSP survey area in \citet{2018PASJ...70S..12T}. 
We also use 151 (60 in $H$-ATLAS) SDSS QSOs at $3.3<z<4.2$ 
studied in \citet{2018PASJ...70S..32U} 
which shows that only two out of the 151 QSOs reside in the protoclusters selected by \citet{2018PASJ...70S..12T}. 

To obtain the average flux as that from a single object, 
it is ideal to stack only isolated sources 
and measure flux on PSF matched images with a sufficiently large aperture;
however, the PSFs of the images used here are extremely large to use such a robust method. 
We here stack them without any smoothing.  
The fluxes measured with $2\times$ FWHM 
of the PSF aperture diameter on each image 
are adopted as approximate estimates of the total fluxes. 
Notably, contaminations from other sources 
are not likely negligible even for $WISE$ and $Herschel$, 
and are considerably large for $IRAS$, $AKARI$, and $Planck$ images. 

As the $g,~r,~i,~z~\&~y$-band flux values and errors, 
the median and standard deviation 
of the Cmodel fluxes of them from the HSC-SSP 
catalog pdr1 \citep{2018PASJ...70S...4A} are used. 

\begin{figure*}
\includegraphics[width=170mm]{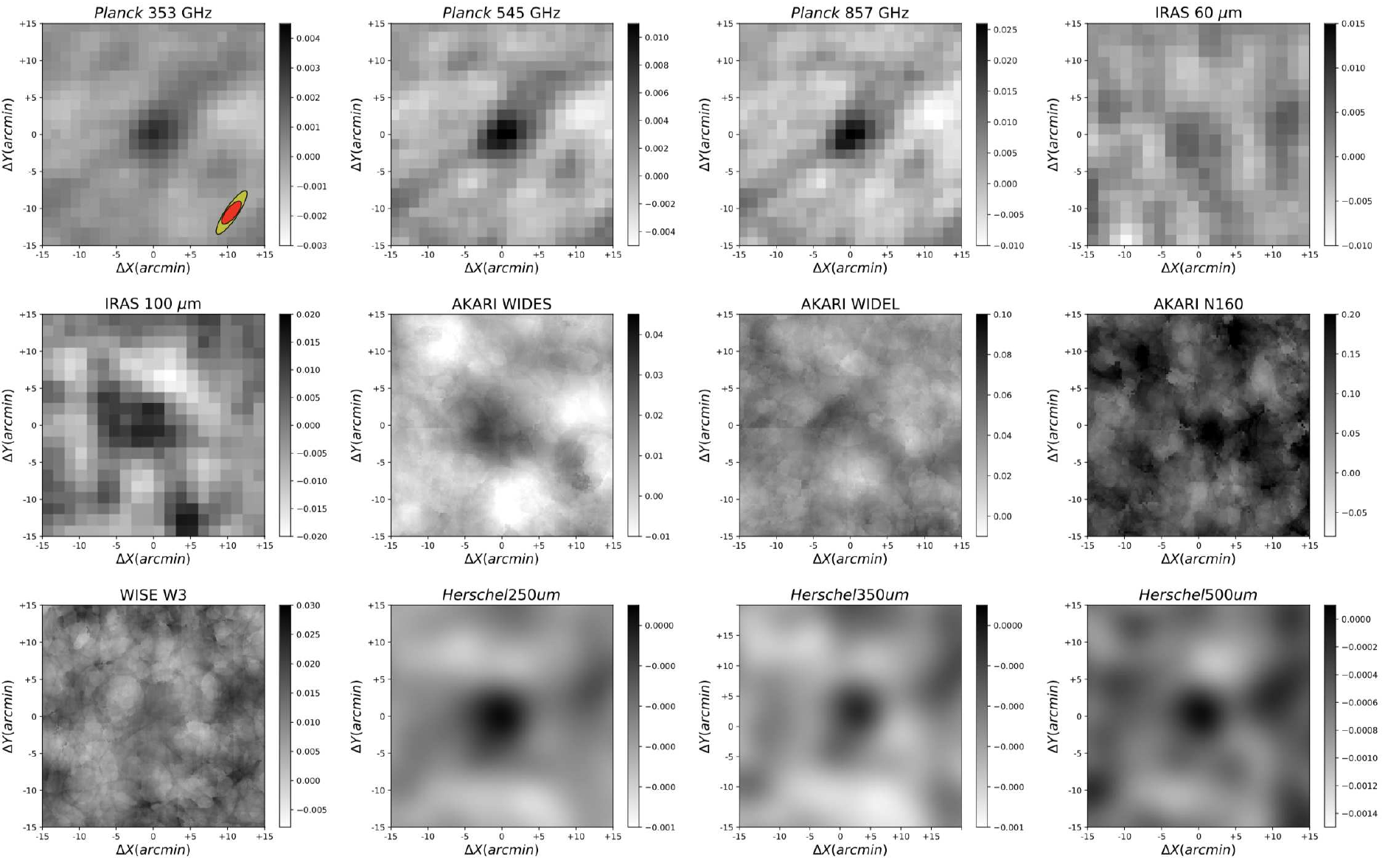}
\caption{ Stacked images of the protoclusters 
in {\it Planck} 353, 545, and 857 GHz; {\it IRAS} 60 and 100 $\mu$m;
{\it AKARI} {\it WIDES, WIDEL}, and {N160}; {\it WISE} {\it W3};,  and {\it Hershcel} 250, 350, and 500 $\mu$m 
from the {\it top left} to {\it bottom right}.
All the images are 30 arcmin by side. 
The ellipses on the 353-GHz images show the average beam profiles described in Section 3.3.
In the case of a point source, 50\% and 90\% of fluxes are included within the red and yellow-filled ellipses, respectively. 
}
\label{fig:stamps} 
\addtocounter{figure}{-1}
\includegraphics[width=170mm]{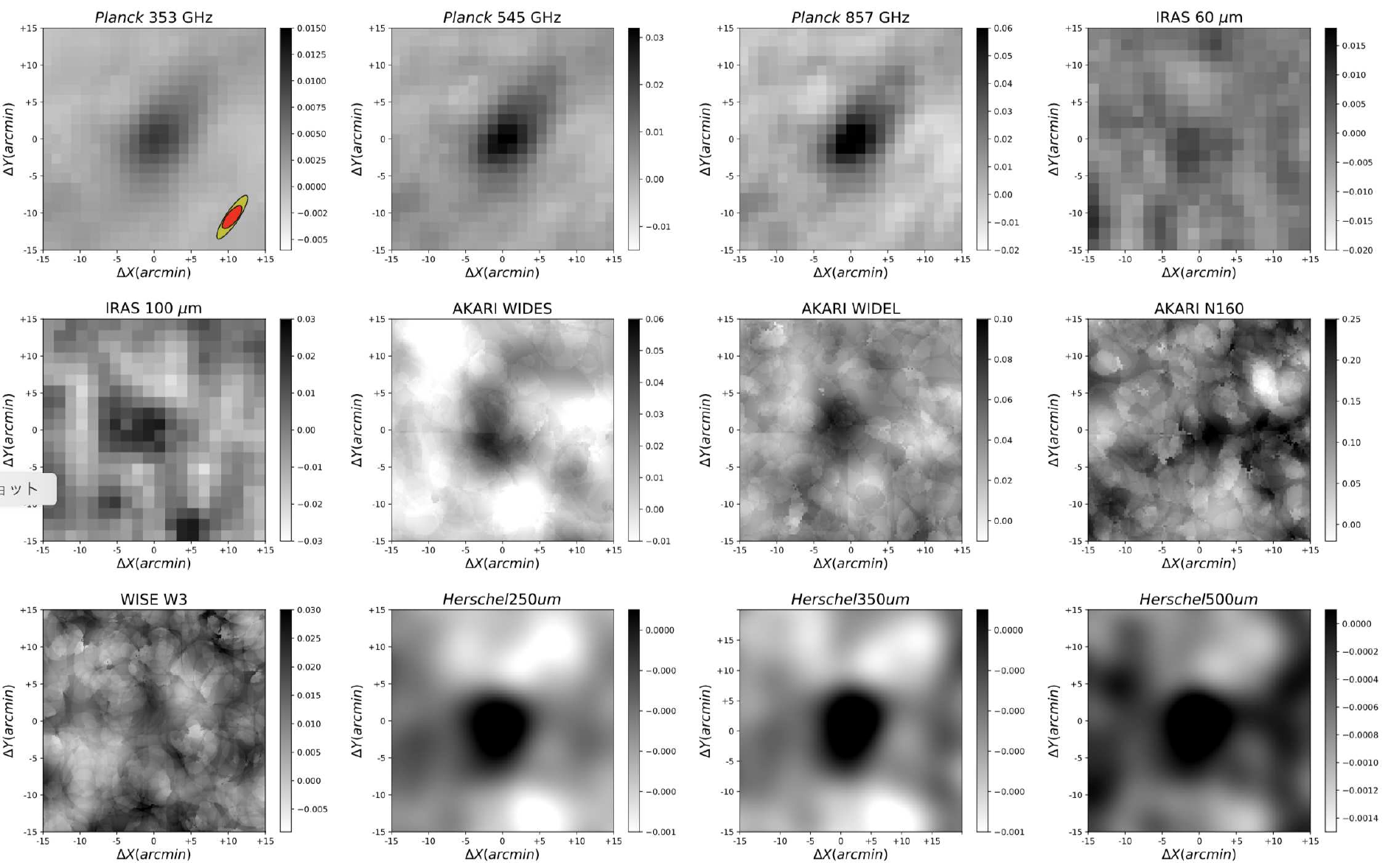}
\caption{ - continues. The stacked images of the brighter-half protoclusters. 	}
\end{figure*}

\begin{figure*}
\addtocounter{figure}{-1}
\includegraphics[width=170mm]{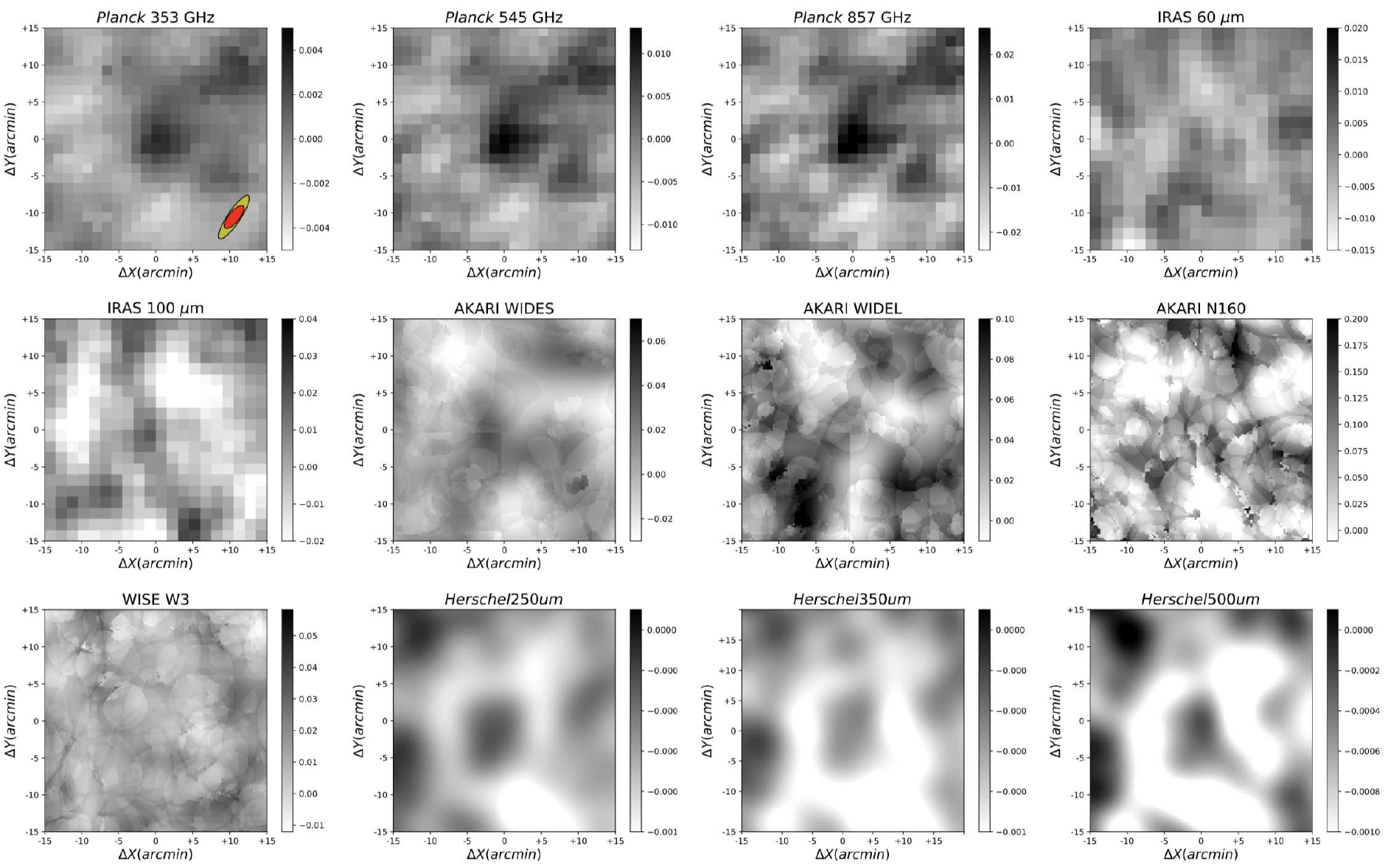}
\caption{ - continues. The stacked images of $5\sigma$ overdensity protoclusters. } \label{fig:stamps5s}	\end{figure*}

\begin{deluxetable*}{lcccccccc}
\tablecaption{The flux densities of HSC SSP $g$-dropout protoclusters\label{tab:flux2}}
\tablewidth{0pt}
\tablehead{
\multicolumn{1}{l}{} &  \multicolumn{2}{c}{{\it IRAS}} &  \multicolumn{3}{c}{{\it Planck}}  &  \multicolumn{3}{c}{{\it Herschel}}\\
\colhead{Sample} & 	\colhead{60~$\mu$m} & \colhead{100~$\mu$m } & \colhead{857 GHz } & \colhead{545 GHz } & \colhead{353 GHz} & \colhead{250~$\mu$m} & \colhead{350~$\mu$m} & \colhead{500~$\mu$m}  \\
\colhead{} & 	\colhead{} & \colhead{} & \colhead{(350~$\mu$m)} & \colhead{(540~$\mu$m)} & \colhead{(840~$\mu$m)} & \colhead{}&\colhead{} &\colhead{} \\
\colhead{} & \colhead{(mJy)} & \colhead{(mJy)} & \colhead{(mJy)} & 	\colhead{(mJy)} & \colhead{ (mJy)} & \colhead{(mJy) } & \colhead{(mJy)} & \colhead{(mJy)} 
}
\decimalcolnumbers
\startdata
All & $6.2\pm2.5$ & $16.8\pm5.5$ & $24.7\pm4.3$ & $11.3\pm2.2$ & $3.7\pm0.8$ & $53.8\pm13.3$ & $24.0\pm8.4$ & $12.6\pm5.3$\\
Brighter-half & $8.9\pm3.6$ & $23.3\pm9.4$ & $69.6\pm4.2$ & $33.8\pm2.2$ & $11.6\pm0.9$& $96.2\pm19.8$ & $58.2\pm11.6$ & $35.4\pm9.1$\\
$5\sigma$ overdensity & $<7.9$ & $14.7\pm6.7$ & $27.0\pm6.3$ & $11.8\pm3.2$ & $3.5\pm1.2$& $45.3\pm14.3$ & $<26.9$ & $<15.1$\\
\enddata 
\tablecomments{Flux values are measured in 4-arcmin diameter aperture. Each flux and error are the average and standard deviation of a thousand times bootstrap resampling.  If they are not detected above 2$\sigma$, we put a 2$\sigma$ value as an upper limit.}
\end{deluxetable*}
\begin{deluxetable*}{lcccccccc}
\tablecaption{ --continues}
\tablenum{1}
\tablewidth{0pt}
\tablehead{
\multicolumn{1}{l}{} &  \multicolumn{4}{c}{{\it AKARI}} &  \multicolumn{4}{c}{{\it WISE}} \\
\colhead{ Sample} & 	\colhead{{\it N60}} & 	\colhead{{\it WIDE-S}} & 	\colhead{{\it WIDE-L}} & 	\colhead{{\it N160}} & 	\colhead{{\it W1}} & 	\colhead{{\it W2}} & 	\colhead{{\it W3}} & 	\colhead{{\it W4}}  \\
\colhead{} & 	\colhead{(65~$\mu$m)} & 	\colhead{(90~$\mu$m)} & 	\colhead{(140~$\mu$m)} & 	\colhead{(160~$\mu$m)} & 	\colhead{(3.4~$\mu$m)} & 	\colhead{(4.6~$\mu$m)} & 	\colhead{(12~$\mu$m)} & 	\colhead{(22~$\mu$m)}  \\
\colhead{} & 	\colhead{(mJy)} & 	\colhead{(mJy)} & 	\colhead{(mJy)} & 	\colhead{(mJy)} & 	\colhead{(mJy)} & 	\colhead{(mJy)} & 	\colhead{(mJy)} & 	\colhead{(mJy)}  
}
\decimalcolnumbers
\startdata
All	   & $<41.5$ & $25.9\pm5.9$ & $<22.7$ 		    & $<70.6$ 			& $<0.82$ & $<0.57$ & $0.40\pm0.13$ & $< 0.57$\\
Brighter-half 		   & $<58.5$ & $43.8\pm8.2$ & $67.1\pm16.8$ & $106.4\pm49.9$ & $<0.98$ & $<0.57$ & $0.54\pm0.18$ & $< 0.86$ \\
$5\sigma$ overdensity & $<77.6$ & $23.2\pm9.7$ & $<40.5$            & $<118.5$	       & $<1.84$ & $<0.92$ & $<0.56$ & $< 1.28$ \\
\enddata
\end{deluxetable*}

\begin{figure*}
\centering
\includegraphics[width=150mm]{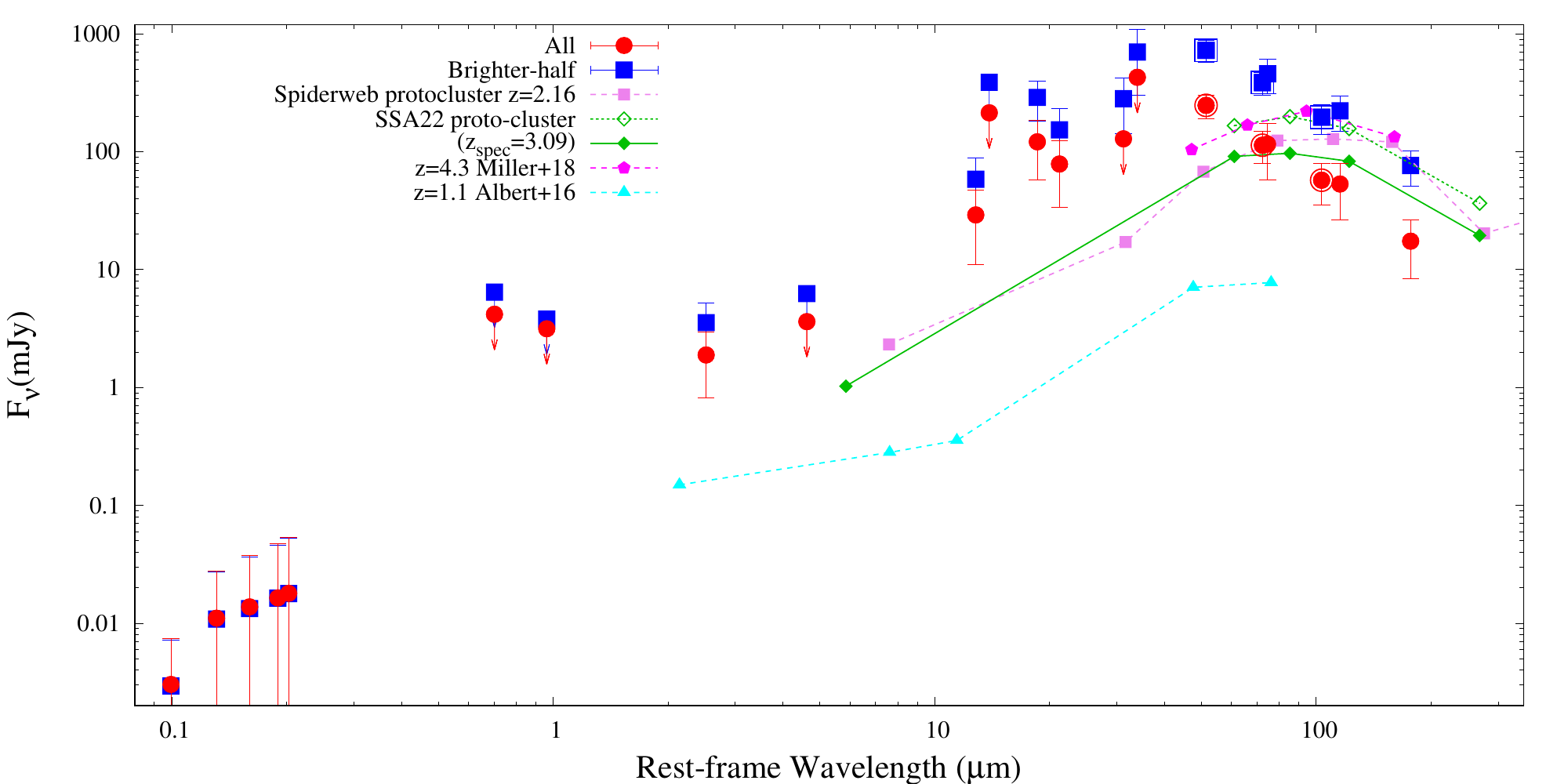}
\caption{
SEDs of the protoclusters at $z=3.8$ and the known protoclusters. 
The red filled circles and blue filled squares show all and the brighter-half protoclusters. 
The encircled symbols show that measured with $H$-ATLAS. 
We show the sum of the IR sources 
in the known protoclusters within the same physical size (8 arcmin $\approx3.4$ physical Mpc diameter). 
We scale all of them to be at $z=3.8$. 
The violet filled squares with a dashed line show the Spiderweb protocluster at $z=2.16$ 
based on \citet{2014A&A...570A..55D}.
The green open diamonds with a dotted line is the SSA22 protocluster at $z\approx3.09$ 
based on \citet{2016MNRAS.460.3861K} and \citet{2014MNRAS.440.3462U}. 
The green filled diamonds with a solid line is that limit to $z_{\rm spec}\approx3.09$ \citep{2018PASJ...70...65U}. 
The data point at 6 $\mu$m is the sum of the fluxes of sources at 
$z_{\rm spec}\approx3.09$ in \citet{2018PASJ...70...65U}
measured with {\it Spitzer} MIPS 24 $\mu$m image in archive \citep{2009ApJ...692.1561W}. 
The magenta filled pentagons with a dashed line are a protocluster at $z=4.3$ in \citet{2018Natur.556..469M}.
The cyan filled triangles with a dashed line are a protocluster at $z=1.1$ in \citet{2016ApJ...825...72A}.
}\label{fig:wlbg}\end{figure*}

\section{Result}

Figure \ref{fig:stamps} shows the stacked images of all, 
brighter-half, and 5$\sigma$-overdensity protoclusters.
Table \ref{tab:flux2} summarizes their fluxes measured using a 4-arcmin diameter aperture.
Signals are significantly detected in {\it WISE} $W3$; {\it IRAS} 60 and 100 $\mu$m; {\it AKARI} {\it WIDES} (and {\it WIDEL} and {N160} for the brighter-half); {\it Planck} 353, 545, and 857 GHz, and $Herschel$ 250, 350, and 500 $\mu$m. 
Although the spectroscopic follow-up of the protoclusters remains on-going, 
it is shown that they trace special environments with excess IR emission. 
Fig. \ref{fig:wlbg} shows the SEDs in the total flux obtained by 
multiplying the aperture fluxes using the aperture correction factors found in Section 3.4. 
{\it This is the first time the ``average"  SED of a protocluster is shown}. 

The flux values of all and the 5$\sigma$-overdensity protoclusters are identical 
although 5$\sigma$ is a more reliable overdensity threshold. 
It implies that the 4$\sigma$ selection is as reliable 
as the 5$\sigma$ selection of the protoclusters.
The brighter-half protoclusters are twice brighter in the {\it Planck} than 
all the protoclusters while there is no significant difference in the optical. 
This implies that above the $4\sigma$ overdensity threshold, 
there is no strong correlation between the optical and IR properties on average.  
Our study demonstrates that deep multi-wavelength observations
are necessary to characterize protoclusters. 

Fig. \ref{fig:fh1} and \ref{fig:fh2} in Appendix C
show the stacked images of the 1st and 2nd quartiles from the lowest 
of the flux distribution at 857 GHz (Fig. \ref{fig:fluxdist}).
The 2nd quartile is marginally detected while the 1st quartile 
shows negative detections perhaps because of noise. 
This indicates the possibility of the artificial signal on the $Planck$ HFI images. 
%This indicates the possibility that they are artificially caused by the reduction procedure of the $Planck$ HFI images. 
However, because the $Herschel$ and $Planck$ results match quite well (Fig. \ref{fig:hpai}), 
they should be negligible. 

In the followings, our results are compared to the known protoclusters at various redshifts
and various populations at $z\sim4$. 

\begin{figure*}
\centering
\includegraphics[width=150mm]{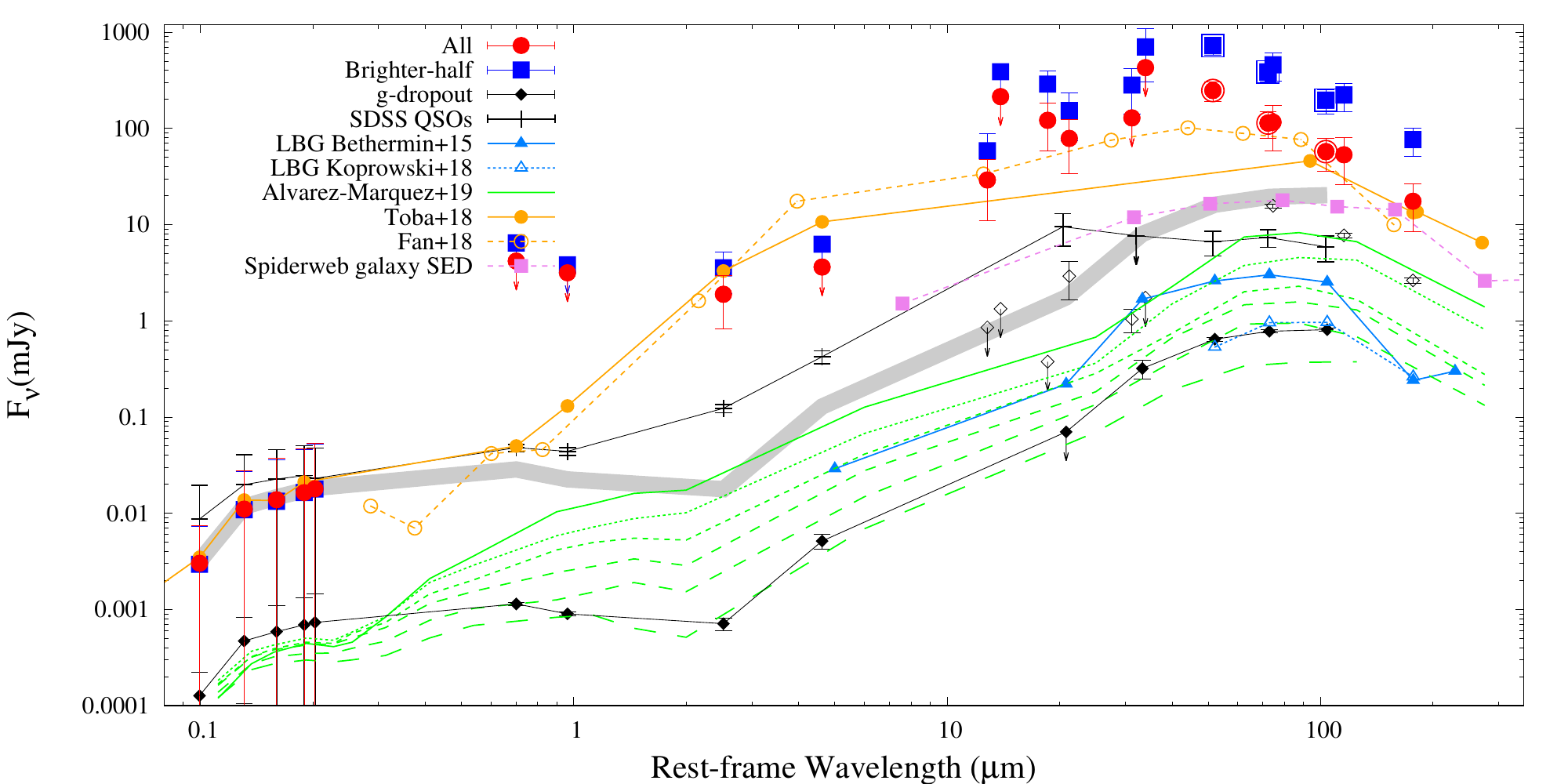}
\caption{Similar to Fig.\ref{fig:wlbg} but we compare with various galaxies at $z\sim4$. 
The black filled diamonds with a solid line show the average flux of a $g$-dropout galaxy measured in Section 3.6. 
Those measured in $Planck$, $AKARI$, and $IRAS$ are shown with open symbols 
since they are severely contaminated by their neighbor sources.  
The black crosses with a solid line show the average flux of a SDSS QSO at $z\sim4$ measured in Section 3.6. 
The light blue filled triangles with a solid line and open triangles with a dot line show 
the average flux of a typical SFG at $z\sim4$ in \cite{2015A&A...573A.113B} and \citet{2018MNRAS.479.4355K}, 
respectively. 
The green curves show the average fluxes of LBGs at $z\sim3$ 
split by stellar mass in \citet{2019A&A...630A.153A}, scaled at $z=3.8$. 
The orange filled circles with a solid line and open circles with a dashed line 
show infrared luminous DOGs at $z=3.7$ in \citet{2018ApJ...857...31T}
and at $z=4.6$ in \citet{2018ApJ...854..157F} scaled at $z=3.8$, respectively. 
The violet filled squares with a dashed line
show the SED of the HzRG in the Spiderweb protocluster \citep{2014A&A...570A..55D}.
The gray shaded region shows the SEDs of the $g$-dropout galaxies multiplied by 20 to 30.
}\label{fig:wlbg2}\end{figure*}

\subsection{Comparison to the known protoclusters}

First, we compare our results with the known protoclusters 
deeply observed in far-infrared (FIR) in Fig. \ref{fig:wlbg}. 
We show the sum of the fluxes of the IR sources
at the same physical size (8 arcmin $\approx$ 3.4 physical Mpc diameter)
of the Spiderweb (MRC1138-262) protocluster 
at $z=2.16$ \citep{2014A&A...570A..55D}, 
the SSA22 protocluster at $z=3.09$ 
\citep{2009ApJ...692.1561W,2016MNRAS.460.3861K,2018PASJ...70...65U}, 
a protocluster at $z=4.3$ reported in \citet{2018Natur.556..469M}
and a massive cluster at $z=1.1$ in \citet{2016ApJ...825...72A}.
The IR sources in the protoclusters at $z>2$ are selected as sub-mm sources 
while those in \citet{2016ApJ...825...72A} are selected at 100 $\mu$m. 
The fluxes of the objects 
with spectroscopic redshifts $z_{\rm spec}$ and/or photometric redshifts $z_{\rm phot}$ similar to the protoclusters
are summed. 
The sum of the purely spectroscopically confirmed sources for the SSA22 protocluster is also shown. 
Their fluxes are scaled to be at $z=3.8$ by multiplying
with (1+3.8)/$(1+z_{\rm at~known~(proto)cluster})\times (D_{\rm L~at~known~(proto)cluster})^2 / (D_{\rm L~at~z=3.8})^2$ 
where $D_{\rm L}$ is the luminosity distance.
Note that only the sources brighter than ultra-luminous infrared galaxies (ULIRGs) 
or hyper-luminous infrared galaxies (HyLIRGs) are counted in the known protoclusters at $z>2$ 
($\gsim4$ mJy at 850 $\mu$m for Spiderweb and $\gsim0.4$ mJy at 1.1 mm for SSA22).
In the case of the SSA22 protocluster, the detections 
of the $X$-ray selected AGNs at $z_{\rm spec}\approx3.09$
\citep{2009ApJ...691..687L,2015ApJ...799...38K} in 24 $\mu$m are also checked. 
All the 24 $\mu$m detected AGNs 
are already included in the sum shown in Fig. \ref{fig:wlbg}.

At $>100~\mu$m in the rest-frame, the flux from the protoclusters  at $z\sim3.8$
and the known protoclusters at $z=2-4$ do not differ in order. 
Amazingly, the Spiderweb and SSA22 protoclusters 
are just as luminous as the typical massive protoclusters at $z\sim3.8$
though they have been believed to be the most prominent structures at $z=2-3$. 
In addition, the SEDs of the Spiderweb and SSA22 protoclusters 
more rapidly decrease at $<100~\mu$m than those of the protoclusters at $z\sim3.8$.
Although only the bright sources in the known protoclusters are summed, 
this tendency may not appreciably change by adding the fluxes 
from IR faint sources optically detected (Section 4.2 and 5). 
Our results imply that the Spiderweb and SSA22 protoclusters may not be particularly special protoclusters in the IR,  
and/or the typical IR luminosities and SEDs of the protoclusters 
have changed drastically between $z=2$ and 4. 

\subsection{Comparison to LBGs, SDSS QSOs and infrared luminous DOGs at $z\sim4$}

Next, we compare the SEDs of the protoclusters
with those of typical SFGs (\citealt{2015A&A...573A.113B,2018MNRAS.479.4355K} 
and $g$-dropout galaxies from the HSC-SSP survey)
and IR luminous DOGs \citep{2018ApJ...854..157F,2018ApJ...857...31T} 
at $z\sim4$ in Fig. \ref{fig:wlbg2}. 

The blue filled and open triangles in Fig. \ref{fig:wlbg2} show the average SEDs of typical SFGs at $z\sim4$ 
measured by stacking analysis in \citet{2015A&A...573A.113B} and \citet{2018MNRAS.479.4355K}. 
The green curves show LBGs at $z=3$ split by stellar mass in \citet{2019A&A...630A.153A}, scaled at $z=3.8$. 
The black diamonds show the average SED of the $g$-dropout galaxies with an $i\leq25.0$ mag 
in the HSC-SSP survey obtained in Section 3.6. 
The $Planck$, $AKARI$, and $IRAS$ fluxes shown with open symbols deviate from $Herschel$ and $WISE$. 
These may be contaminated by surrounding $g$-dropout galaxies 
as well as some unknown protoclusters because of low spatial resolution. 
The average SED of the $g$-dropout galaxies 
selected from the HSC-SSP survey matches well 
\citet{2018MNRAS.479.4355K} and \citet{2019A&A...630A.153A} 
whose sample selections are similar to ours.
\citet{2015A&A...573A.113B} is biased to more massive objects
and there is no wonder that it does not match our results. 
The gray shaded region shows the SED of $g$-dropout galaxies multiplied by 20 to 30 which is
the expected number of $g$-dropout galaxies with an $i\leq25$ mag in a protocluster. 
The protoclusters are not only several tens of times 
brighter than typical SFGs but they have SEDs 
with greater warm/hot dust component compared to those of typical SFGs at $z\sim4$.  
From the aforementioned, we argue that the IR SEDs
of the protoclusters cannot be explained 
by only multiplying typical SFGs at $z\sim4$. 

The dust torus of an AGN are luminous in the mid to FIR; 
however, at least, SDSS QSOs or optically luminous QSOs
are not found at the HSC-SSP protoclusters in general \citep{2018PASJ...70S..32U}. 
Our results suggests that there are overdensities of IR sources that 
cannot be selected by $g$-dropout selection 
and/or $g$-dropout galaxies in the protoclusters have special properties 
such as AGN-dominated DOGs; 
The extremely IR luminous DOGs detectable with $WISE$ 
at $z=3.7$ \citep{2018ApJ...857...31T} 
and at $z=4.6$ \citep{2018ApJ...854..157F} are shown in Fig. \ref{fig:wlbg2}. 
\citet{2018ApJ...857...31T} is shown without any scaling 
while \citet{2018ApJ...854..157F} is plotted after scaling the flux at $z=3.8$.
Interestingly, the SEDs of the protoclusters quite resemble those of the luminous DOGs. 
They report that these DOGs have IR SEDs dominated by dust emission from AGN tori. 
Because the AGN emission dominates $\sim50$ \% of the total flux even 
at $\sim200~\mu$m in the rest-frame, despite the huge total IR luminosity,  
these DOGs have a SFR of only $\sim480$ and 1300 $M_{\odot}$ yr$^{-1}$
for \citet{2018ApJ...854..157F} and \citet{2018ApJ...857...31T}, respectively.
This result implies that dust emission from AGN tori
and moreover a single object such as IR luminous DOGs 
can dominate the fluxes of the protoclusters. 

We discuss the breakdown of the IR emission from the protoclusters 
by fitting them with models as in the next section. 

\begin{figure*}\begin{center}
\includegraphics[width=87mm]{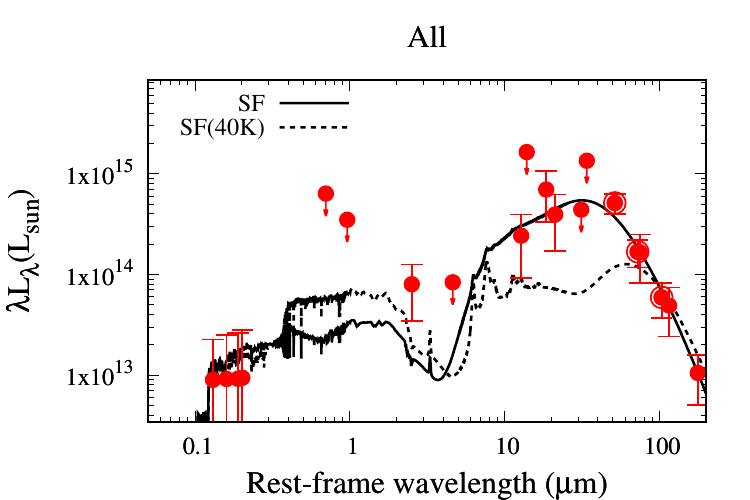}
\includegraphics[width=87mm]{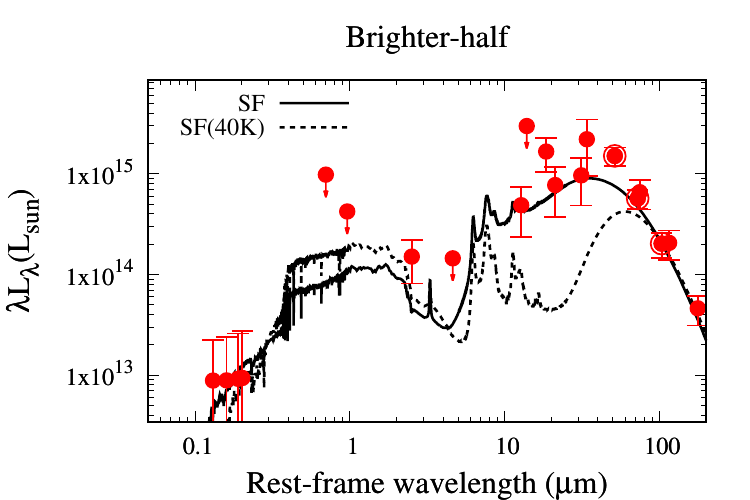}
\includegraphics[width=87mm]{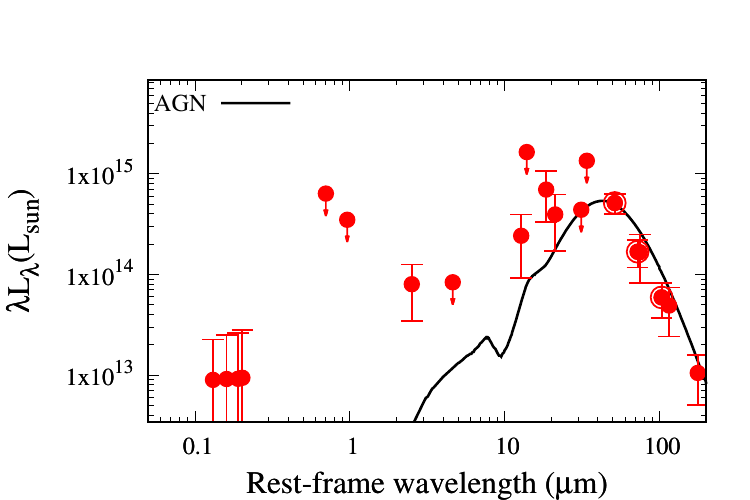}
\includegraphics[width=87mm]{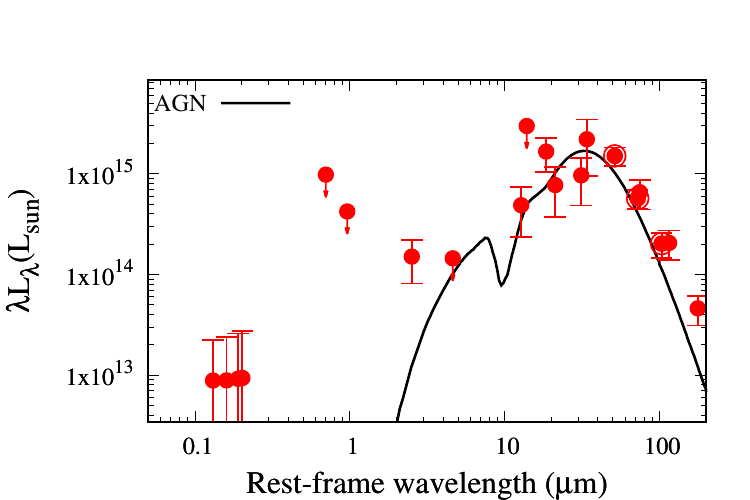}
\includegraphics[width=87mm]{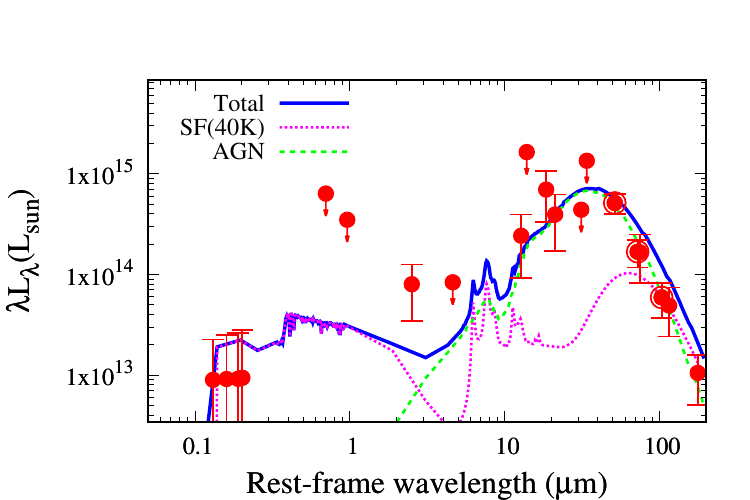}
\includegraphics[width=87mm]{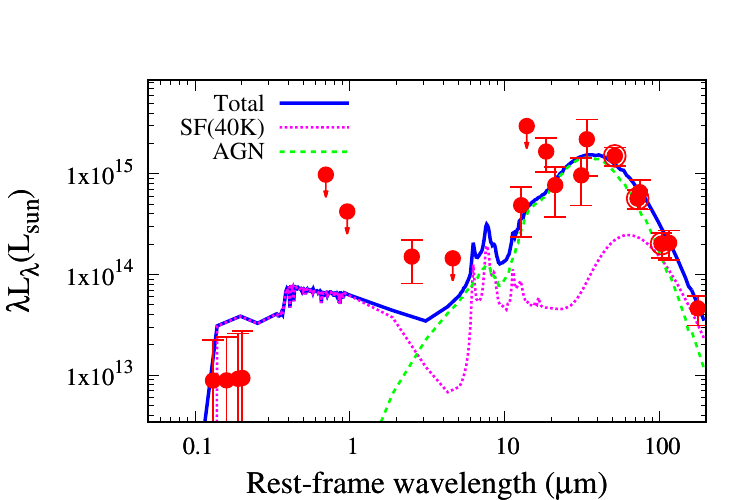}
\end{center}
\caption{{\it Left:} Best-fit SEDs for all the protoclusters. The red filled circles show the observed fluxes. 
{\it Left top:} The black solid curve shows the best-fit SED found with {\sf MAGPHYS}. 
The black dashed curve shows the best-fit SED found with {\sf MAGPHYS} limit to $T_{dust}\approx 40~K$. 
{\it Left middle:} The black solid line shows the best-fit SED for AGN SED models.
{\it Left bottom:} The best-fit model for composite SED models 
of a {\sf MAGPHYS} and an AGN SED model.
The blue solid, magenta dotted, and green dashed lines 
show the total, star-forming, and AGN components of the best-fit SEDs, respectively. 
{\it Right:} Similar to the {\it left} panels but for the brighter-half protoclusters.}
\label{fig:sedfits1}
\end{figure*}
\setcounter{figure}{8}
\begin{figure}
\includegraphics[width=87mm]{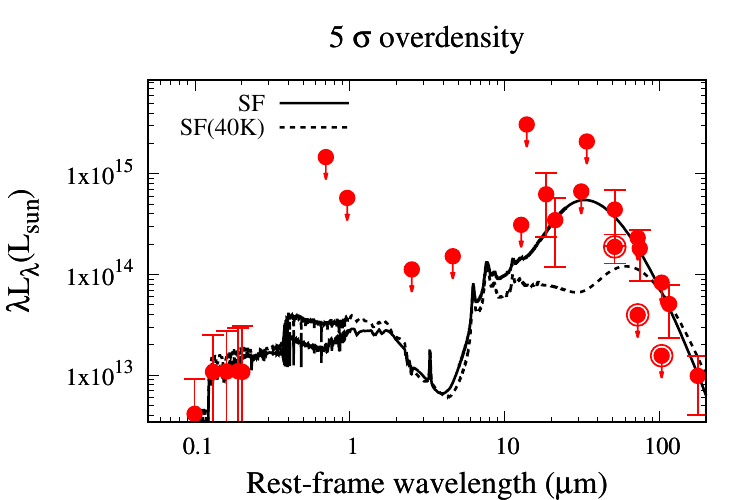}
\includegraphics[width=87mm]{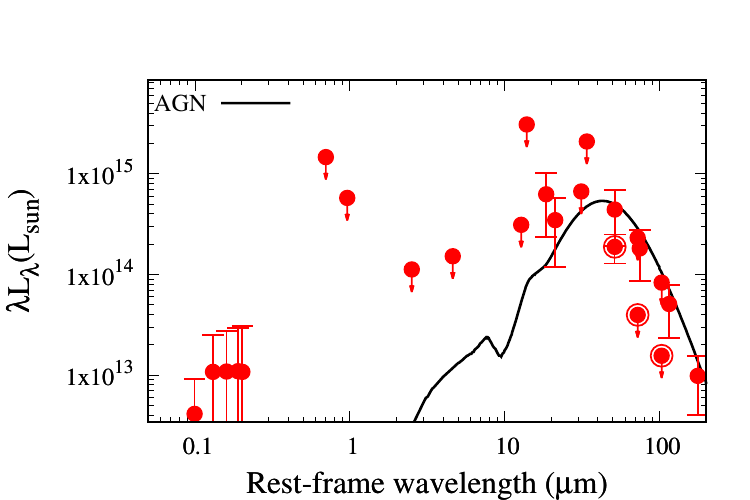}
\includegraphics[width=87mm]{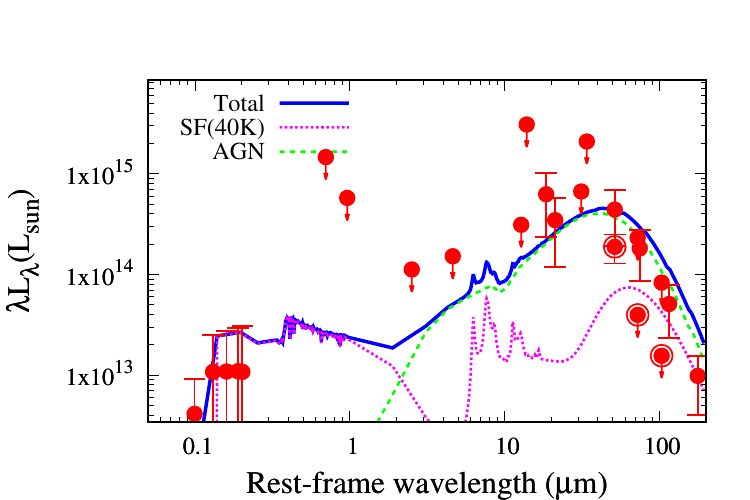}
\caption{ --continues. Similar to the above figures of Fig. \ref{fig:sedfits1} but for the 5$\sigma$ overdensity protoclusters.}

\end{figure}

\begin{deluxetable*}{ccccc}
\tablecaption{$\chi^2/\nu$ values for SED models\label{tab:lirchi2}}
\tablewidth{0pt}
\tablehead{
\colhead{Sample} & \colhead{{\sf MAGPHYS}} & \colhead{{\sf MAGPHYS} ($T_{dust}=40 K$)} & \colhead{AGN } & \colhead{{\sf MAGPHYS}+AGN}
}
\decimalcolnumbers
\startdata
All	            &  0.53   &   2.50  & 0.74 &  0.62\\
Brighter-half	     &  0.76  &   3.60  & 1.50 &  0.91\\
$>5\sigma$ overdensity &  0.55  &   1.46  & 0.55  &  0.46\\
\enddata
\end{deluxetable*}

\begin{deluxetable*}{ccccccc}
\tablecaption{Best fit SED model (model: {\sf MAGPHYS} ($T_{dust}=40 K$) + AGN)\label{tab:lir}}
\tablewidth{0pt}
\tablehead{
\colhead{Sample} & \colhead{$L_{FIR, SB}$} & \colhead{$L_{FIR, AGN}$} & \colhead{$L_{FIR, total}$} & \colhead{$L_{FIR, SB}/L_{FIR, AGN}$} & \colhead{ SFR} & \colhead{$A_V$} \\
 \colhead{}   &  \colhead{($10^{13}~L_{\odot}$)}&  \colhead{($10^{13}~L_{\odot}$)} &  \colhead{($10^{13}~L_{\odot}$)}  & \colhead{} &  \colhead{($10^3~M_{\odot}$ yr$^{-1}$ )}   &  \colhead{}	 
}
\decimalcolnumbers
\startdata
All	            & $1.3_{-1.0}^{+1.6}$ & $3.7_{-2.0}^{+1.8}$ & $5.1_{-2.5}^{+2.5}$     & $0.4^{+0.7}_{-0.2}$ & $2.1^{+6.3}_{-1.7}$ & $0.5^{+1.0}_{-0.2}$\\
Brighter-half		     & $3.2_{-1.0}^{+4.5}$ & $11.1_{-3.6}^{+3.5}$ & $14.2_{-4.5}^{+5.8}$ & $0.3^{+0.5}_{-0.0}$ & $1.9^{+3.9}_{-0.9}$ & $1.0^{+0.0}_{-0.4}$\\
$>5\sigma$ overdensity & $0.9_{-0.8}^{+1.4}$ & $2.8_{-2.3}^{+1.7}$ & $3.7_{-2.9}^{+2.1}$      & $0.3^{+1.8}_{-0.3}$ & $1.1^{+2.9}_{-1.0}$ & $0.3^{+1.8}_{-0.0}$\\
\enddata
\end{deluxetable*}

\begin{deluxetable*}{ccccc}
\tablecaption{Best fit SED model (model: {\sf MAGPHYS})\label{tab:lirmagphys}}
\tablewidth{0pt}
\tablehead{
\colhead{Sample} & \colhead{$T_{dust}$} & \colhead{$L_{dust}$} & \colhead{SFR} & \colhead{$A_V$}  \\
 \colhead{}   &  \colhead{} &  \colhead{($10^{13}~L_{\odot}$)}   &  \colhead{($10^3~M_{\odot}$ yr$^{-1}$ )}   &  \colhead{}	 
}
\decimalcolnumbers
\startdata
All	            & $71_{-2}^{+4}$ & $19.3_{-4.2}^{+0.6}$   & $16.3_{-7.8}^{+1.0}$   & $1.7^{+1.6}_{-0.1}$\\
Brighter-half	     & $74_{-12}^{+0}$ & $48.7_{-7.1}^{+0.7}$   & $43.4_{-8.2}^{+2.9}$ & $1.8^{+1.9}_{-0.0}$\\
$>5\sigma$ overdensity & $75_{-20}^{+0}$ & $16.0_{-11.7}^{+8.2}$   & $13.4_{-10.1}^{+6.9}$   & $1.7^{+3.4}_{-0.4}$\\
\enddata
\end{deluxetable*}

\section{Discussion}
\label{sec:discussion}
\subsection{Origin of the IR emission}
\label{sec:discussionsed}

First, the sum of the IR fluxes of the $g$-dropout galaxies of the protoclusters, 
estimated by multiplying the average flux of a $g$-dropout galaxy 
with the number excess of $g$-dropout galaxies ($N=20\sim30$), 
is only a third and a tenth of the flux of all and the brighter-half protoclusters, respectively. 
In addition, the average SED of the $g$-dropout galaxies is different from that of the protoclusters.
Therefore the $g$-dropout galaxies are not sufficient to explain the whole IR flux of the protoclusters. 
Obscured AGNs are plausible origin of the MIR excess \citep{2018ApJ...854..157F,2018ApJ...857...31T}. 
According to the previous studies of the protoclusters shown in Fig. \ref{fig:wlbg}, 
several SMGs comprise the remaining greater portion of the IR luminosity of the protoclusters. 
It is also reported that such sources found by single-dish telescopes 
are resolved into multiple SMGs by ALMA (e.g., \citealt{2018PASJ...70...65U}).

We discuss the origin of the IR emission of the protoclusters 
by fitting the observed SEDs with the model SEDs. 
The major components of a SED from $\approx0.01$ to $200~\mu$m in the rest-frame 
are stellar emission, and emission from dust heated by young stars and AGN torus 
(Here we ignore other AGN components since we focus on the dust emission in the IR).
Therefore, the SED fitting is performed by using the models of 
(1) stellar emission and emission from dust heated by stars, 
(2) AGN torus, and (3) their combination. 

We adopt the SED models from {\sf MAGPHYS} \citep{2008MNRAS.388.1595D}
which generates the SED models via the combinations 
of stellar light and emission from dust heated by young stars. 
{\sf MAGPHYS} describes the UV to IR SEDs with the consistency 
of the absorbed UV light and that re-emitted in IR.
The dust in {\sf MAGPHYS} consists of that in stellar birth clouds 
and ambient inter-stellar medium.
The former represents hot ($130-250~K$) and warm ($30-60~K$) dust components,
and the latter represents a cold ($15-25~K$) dust component.
Notably, {\sf MAGPHYS} can generate a model containing 
a large warm/hot dust component which can also originate in an AGN torus.
{\sf MAGPHYS} is a comprehensive package 
generating various SED models 
and fitting an observed SED with SED models. 
When combining {\sf MAGPHYS} models with AGN models, 
we extract $\approx4800$ models with solar metallicity 
and dust temperature $T_{dust}\approx40~K$ 
(for SFGs at $z\sim4$ in \citealt{2018MNRAS.479.4355K}) for simplification. 

For SED models of AGNs, 
we adopt the model library by \citet{2015A&A...583A.120S}. 
Their models are parameterized by viewing angle, 
inner radius of the dusty torus $R$, cloud volume filling factor $V_c$, 
optical depth (in $V$-band) of the individual clouds $A_c$ 
and the optical depth (in $V$-band) of the disk mid-plane $A_d$. 
Because the images are stacked,
we use the average of the model SEDs with the same parameters 
but different viewing angles. 
The Lyman forest absorption at $\lambda<1216$~\AA~
in the rest-frame is manually added on AGN models following \citet{1995ApJ...441...18M}. 

Then we fit the observed SEDs in case (1) to (3)
using a standard $\chi^2$ minimization procedure. 
In case (3), we fit the observed SEDs with the models combining SFG and AGN models with free ratio. 
Fig. \ref{fig:sedfits1} shows the best-fit SED models 
and Table \ref{tab:lirchi2} shows their $\chi^2/\nu$ values. 
The $\chi^2/\nu$ values minimize for cases (1) and (3). 
The best-fit SED parameters are shown in Table \ref{tab:lir} 
and \ref{tab:lirmagphys}. 
Though several works suggested high dust temperatures for SFGs at high redshift
(e.g., \citealt{2012ApJ...760....6M,2015A&A...573A.113B, 2016ApJ...833...72B, 2017ApJ...847...21F, 2019MNRAS.489.1397L}),  
the best-fit models of {\sf MAGPHYS} of protoclusters have dust temperatures $T_{dust}\sim70~K$, 
which are exceptionally higher than that of a typical SFG at $z\sim4$, $T_{dust}\sim40~K$ \citep{2018MNRAS.479.4355K}. 
Such high $T_{dust}$ models describe SFGs with a very high specific SFR $\gsim10$ Gyr$^{-1}$. 
In the case of the composite models, $70\sim80~\%$ of the total FIR ($8-1000~\mu$m)  luminosities
originate in AGNs. 
Briefly, the total FIR luminosity of all and the brighter-half protoclusters 
is $5.1_{-2.5}^{+2.5}$ and $14.2_{-4.5}^{+5.8}\times10^{13}~L_{\odot}$, respectively 
for the best-fit {\sf MAGPHYS}+AGN models
and $19.3_{-4.2}^{+0.6}$ and $48.7_{-7.1}^{+0.7}\times10^{13}~L_{\odot}$, respectively,
for the best-fit {\sf MAGPHYS} models. 
The total SFR of all and the brighter-half protoclusters 
is $2.1_{-1.7}^{+6.3}$ and $1.9_{-0.9}^{+3.9}$ $\times10^{3}~M_{\odot}$ yr$^{-1}$, respectively, 
for the best-fit {\sf MAGPHYS}+AGN models 
and $16.3_{-7.8}^{+1.0}$ and $43.4_{-8.2}^{+2.9}$ $\times10^{3}~M_{\odot}$ yr$^{-1}$, respectively, 
for the best-fit {\sf MAGPHYS} models. 

\begin{figure*}
\centering
\includegraphics[width=87mm]{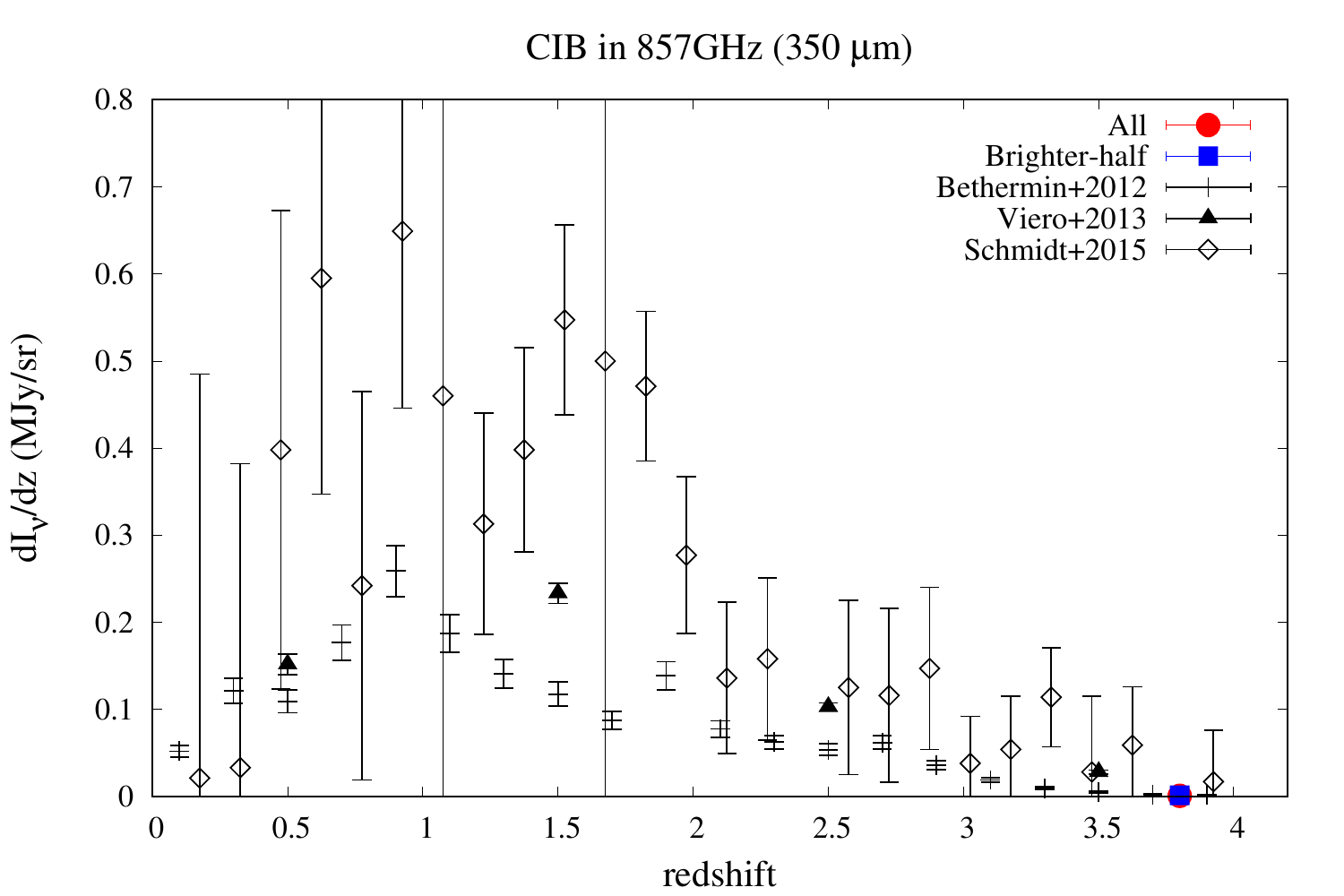}
\includegraphics[width=87mm]{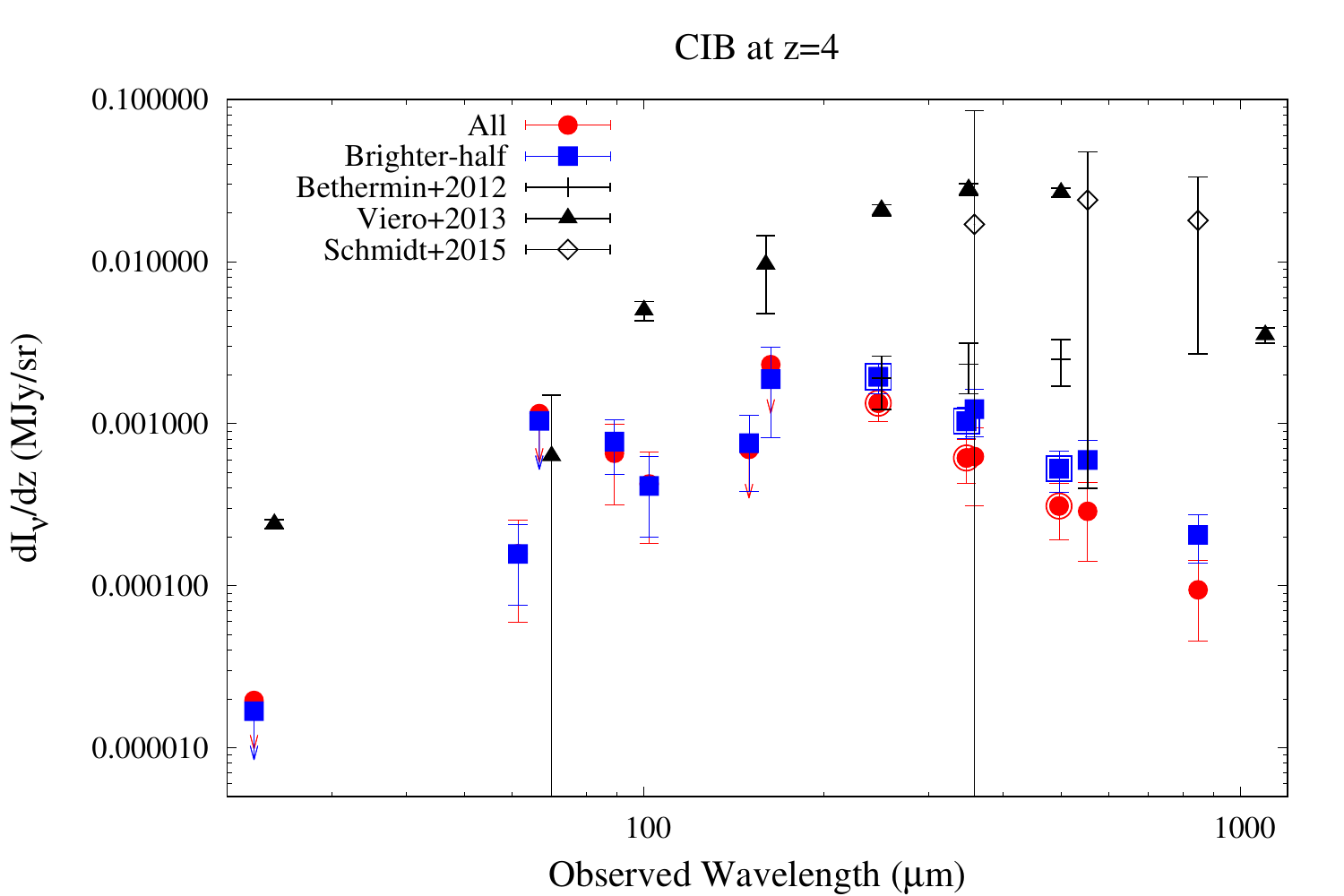}
\caption{
{\it Left:} CIB level redshift distribution in 857 GHz (350 $\mu$m). 
The red filled circle and blue filled square show the CIB 
from all and the brighter-half protoclusters, respectively. 
The black crosses, triangles, and diamonds show the CIB mean levels 
in \citet{2012A&A...542A..58B}, \citet{2013ApJ...779...32V}, and \citet{2015MNRAS.446.2696S}, respectively. 
We show the median value at $3.4<z<4.0$ in Table B1 of \citet{2012A&A...542A..58B}, 
the value at $3.0<z<4.0$ in Table 6 of \citet{2013ApJ...779...32V}, 
and the median value at $3.325<z<4.225$ in Table 6 of \citet{2015MNRAS.446.2696S}. 
{\it Right:} The CIB level at $z=4$. Symbols are the same as {\it left} panel. 
}\label{fig:cib}\end{figure*}

At this point, whether the warm/hot dust emission from the protoclusters
originates in star formation or AGNs cannot be determined by the SED fitting.   
However, given the dust temperature of typical SFGs at $z\sim4$, 
and the presence of luminous QSOs and/or overdensities 
of AGNs in the known protoclusters at $z=2-3$, 
dust emission from AGNs are likely not negligible. 
Further characterization of galaxies in the protoclusters 
e.g., SEDs with higher S/N ratio and line diagnostics for individual sources
will be helpful to distinguish these scenarios. 

\subsection{Contribution of the protoclusters to the CIB at $z\sim4$}

The cosmic infrared background (CIB; \citealt{2005ARA&A..43..727L,2011A&A...536A..18P}) 
is the cumulative infrared emission from all galaxies/AGNs throughout cosmic history 
\citep{2006A&A...451..417D,2014A&A...571A..30P}. 
The redshift evolution of the mean CIB intensity is an important probe 
of the whole star formation history in the Universe. 
The anisotropy of the CIB traces the large scale distribution of DSFGs
\citep{2011Natur.470..510A,2012A&A...542A..58B,2013A&A...557A..66B,2013ApJ...779...32V,2018A&A...614A..39M}. 
Protoclusters should represent the most biased regions of the CIB. 
Here, we discuss the consistency of our results with the CIB anisotropy studies in literatures. 

Fig. \ref{fig:cib} shows the redshift evolution of the CIB intensity 
at 857 GHz ($\approx 350~\mu$m) and the wavelength 
dependence of the CIB intensity at $z\sim4$. 
The average flux of a protocluster is converted into the CIB intensity in MJy/sr by, 
$dI_{\nu}/dz = F_{\nu}~({\rm MJy}) \times N_{\rm pcl}~(deg^{-2}) \times 3282~(sr~deg^{-2})~/~dz$, 
where the number density of the protoclusters $N_{\rm pcl}~(deg^{-2}) = 179/121$ 
and the redshift range $dz\approx0.9$ according 
to the redshift selection function for $g$-dropout galaxies in \citet{2016ApJ...826..114T}.
\citet{2012A&A...542A..58B} and \citet{2013ApJ...779...32V} obtained the CIB intensity
by stacking the $Herschel$ images of the photometric redshift catalogs. 
\citet{2015MNRAS.446.2696S} evaluated the CIB intensity based on the $Planck$ HFI data 
inferring redshift distribution by taking a cross-correlation with SDSS QSOs. 
Note that \citet{2012A&A...542A..58B}  
is only sensitive for the sources with 24 $\mu$m fluxes $>80\mu$Jy 
while \citet{2013ApJ...779...32V} studied sources fainter than \citet{2012A&A...542A..58B}. 
The cross-correlation method 
(e.g., \citealt{2015MNRAS.446.2696S,2018A&A...614A..39M}) 
is sensitive for further faint 
unresolved populations but only covers the $Planck$ HFI bandpath at this point.

The CIB intensity at $z\sim4$ is $0.02\sim0.03$ MJy/sr in 857 GHz or 350 $\mu$m in the literatures 
\citep{2013ApJ...779...32V,2015MNRAS.446.2696S,2018A&A...614A..39M}. 
All protoclusters in this study have a 350 $\mu$m intensity of $0.0006\pm0.0003$ MJy/sr
while that of the brighter-half protoclusters is $0.0012\pm0.0004$ MJy/sr. 
This implies that we should consider the IR luminosity function of the protoclusters 
to properly evaluate the protocluster contribution to the CIB.
Here, we adopt the value evaluated with the brighter-half protoclusters as a lower limit.  
According to \citet{2018A&A...614A..39M} who evaluated the CIB anisotropy 
based on the $Planck$ CIB auto- and cross-power spectra, 
and the CIB and CMB (cosmic microwave background) lensing cross-spectra, 
the dark matter halos contributing the most to the CIB
have a nearly constant  $M_h\approx10^{12.77}~M_{\odot}$ at $1<z<4$. 
According to them, the contribution of dark matter halos with $M_h>10^{13}~M_{\odot}$, 
which is the typical mass of the protoclusters in \citet{2018PASJ...70S..12T},
to the whole CIB is several percent,
although the volume density of the protoclusters at $z\sim4$ is quite small. 
We find that the protoclusters in \citet{2018PASJ...70S..12T} comprise the $\gsim6$ \% of the whole CIB at $z\sim4$,
consistent with \citet{2018A&A...614A..39M}.

At $< 350~\mu$m, 
the contribution of the protoclusters to the entire CIB intensity becomes
larger than that at a longer wavelength. 
This can reflect the true bias of warm dust emission sources such as AGNs and young starburst galaxies 
but note that there are several observational gaps.; 
The previous studies performed using the stacking analysis of the photometric redshift catalogs 
are subject to the selection incompleteness 
because of the survey depth and the photometric redshift selections 
which tend to miss objects with non-galaxy-like SEDs e.g., QSOs. 
The cross-power spectra method (e.g., \citealt{2018A&A...614A..39M}) assumes  
only the typical SED of SFGs. 
\citet{2015MNRAS.446.2696S} used SDSS QSOs as priors, however, 
they do not always represent the regions brightest in the IR (Section 5.4). 

\begin{figure}
\centering
\includegraphics[width=87mm]{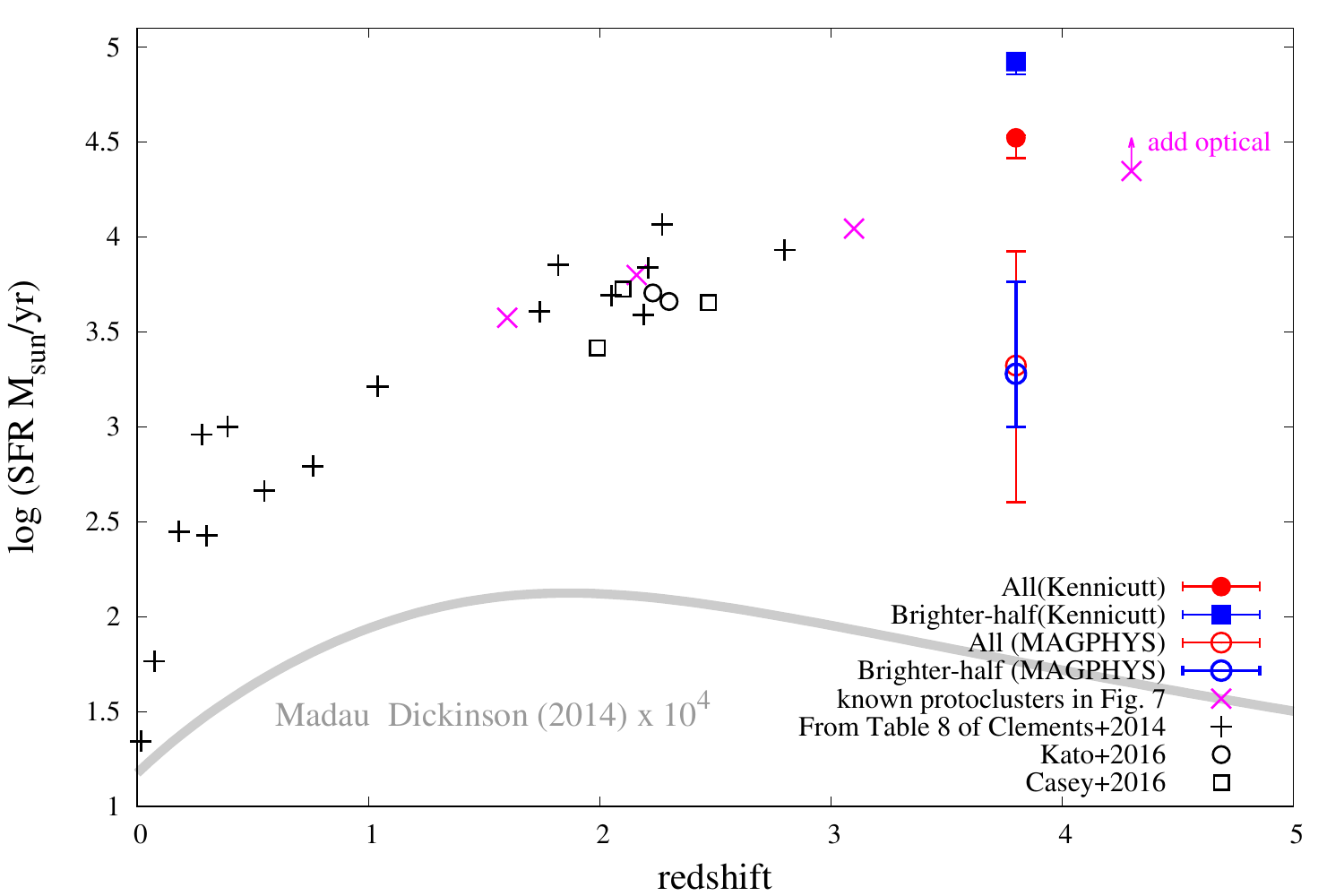}
\caption{Evolution of the SFRs of the protoclusters. 
The red filled circle and blue filled square show the (average) SFR of all and the brighter-half protoclusters 
evaluated by multiplying $L_{FIR}$ in Table \ref{tab:lirmagphys} 
with the convertion factor in \citet{1998ARA&A..36..189K}. 
The open ones are the SFR  in Table \ref{tab:lir} which are subtracted of the contribution of AGNs. 
The black crosses, open circles, and squares show the protoclusers/clusters summarized in  
\citet{2014MNRAS.439.1193C}, \citet{2016MNRAS.460.3861K} and \citet{2016ApJ...824...36C}. 
The magenta crosses show the four protoclusters in Fig. \ref{fig:wlbg}. 
The gray tick curve shows the cosmic SFR density in general field in \citet{2014ARA&A..52..415M} 
multiplied by $10^4$. 
}\label{fig:tsfrredshift}\end{figure}

\subsection{Evolution of SFRs of massive protoclusters}

Figure \ref{fig:tsfrredshift} shows the evolution 
of the SFR of protoclusters/clusters. 
At $z\sim3.8$, we show the total SFRs subtracted of the IR emission from AGNs (Table \ref{tab:lir}) 
and those measured by only multiplying the total FIR luminosities (Table \ref{tab:lirmagphys})  
with a conversion factor in \citet{1998ARA&A..36..189K} 
where SFR $=L_{FIR}\times1.7\times10^{-10}~(M_{\odot}$ yr$^{-1}$). 
Our results are compared to the SFRs of massive protoclusters/clusters.
We refer to the protoclusters/clusters at $0<z<3$ listed in Table 8 of \citet{2014MNRAS.439.1193C} which is 
originally based on \citet{2000A&A...363..933M} (Perseus), \citet{2011MNRAS.412.1187B} (A3112), \citet{2000A&A...361..827F} (A1689), \citet{2009MNRAS.396.1297H} (A1758). 
\citet{2010ApJ...725.1536C} (the Bullet cluster), 
\citet{2006ApJ...649..661G} (Cl0024+16 and MS0451-03), 
\citet{2010MNRAS.405.2623S} (protoclusters around QSOs at $1.7<z<2.8$), 
and clusters of DSFGs selected 
with $Planck$ and $Herschel$ in \citet{2014MNRAS.439.1193C};
2QZ and HS1700 protoclusters in \citet{2016MNRAS.460.3861K};.
the known protoclusters shown in Fig. \ref{fig:wlbg}:, and 
the GOODS-N $z=1.99$ protocluster \citep{2004ApJ...611..725B,2009ApJ...691..560C},
COSMOS $z=2.10$ protocluster \citep{2014ApJ...795L..20Y} and
COSMOS $z=2.47$ protocluster \citep{2015ApJ...808L..33C}
summarized in \citet{2016ApJ...824...36C}. 
\citet{2014MNRAS.439.1193C} measured the total IR luminosities 
of  all the IR sources in protoclusters/clusters by fitting their IR SEDs with a modified black body
with a dust emission index $\beta=2$ and converted them 
to the SFR with the relation in \citet{2003ApJ...586..794B},
which is slightly (10 percent) different from that in \citet{1998ARA&A..36..189K}. 
\citet{2016MNRAS.460.3861K} pre-selected IR sources with photometric redshifts, 
measured the IR luminosities by fitting their IR SEDs with 
a modified black body with a dust emission index $\beta=1.5$, 
and converted them to the SFR with the conversion factor in \citet{1998ARA&A..36..189K}.
The SFRs summarized in \citet{2016ApJ...824...36C} were measured using {\sf MAGPHYS}
or the conversion factor in \citet{1998ARA&A..36..189K}.
For the known protoclusters shown in Fig. \ref{fig:wlbg} (we use the spec-z only flux for the SSA22 protocluster), 
the total FIR luminosities are measured by fitting the total IR SEDs with {\sf MAGPHYS} 
and converting them into the SFR using the conversion factor in \citet{1998ARA&A..36..189K}.
Although there are differences in the methods to obtain total FIR luminosities, 
the SFRs converted using \citet{1998ARA&A..36..189K} or \citet{2003ApJ...586..794B} are evaluated in a similar manner. 

Here, the total SFRs are shown while \citet{2014MNRAS.439.1193C} 
and \citet{2016MNRAS.460.3861K} showed a SFR density-redshift diagram.
Because their considered sizes ($\sim1$ Mpc in physical radius volume)
of high-z protoclusters are smaller than the considered size 
of the protoclusters at $z=3.8$ in this study,  
it is not trivial to calibrate our measurement to their SFR density. 
Notably, according to the empirical source distributions in the known protoclusters 
and our simulation shown in Fig. \ref{fig:sim2}, 
most of the fluxes from the IR sources in the protoclusters 
are likely concentrated within a few arcmin ($\sim1$ Mpc in physical) radius. 
Thus the SFR densities of the protoclusters at the central $\sim1$ Mpc in physical radius volume
may follow the total SFRs well. 
At $z\gsim2$, the literature only considers bright sources more luminous than ULIRGs/HyLIRGs.
In Section 5.1, we found that one third of the flux of a protocluster 
can originate in $g$-dropout galaxies. 
The upward arrow at $z=4.3$ in Fig. \ref{fig:tsfrredshift} 
shows a possible correction because of such galaxies optically selected.

The measured masses of the referred $z\lsim1$ clusters  
are $M_h=6\times10^{14}~M_{\odot}$ to a few $10^{15}~M_{\odot}$ 
while those of protoclusters are approximately $10^{14}~M_{\odot}$. 
Our targets are relatively massive protoclusters 
which will collapse into a halo with halo mass $>10^{14}~M_{\odot}$. 
Therefore, although the selection techniques are not uniform, 
Fig. \ref{fig:tsfrredshift} shows the evolution of the most massive clusters today. 
While the cosmic SFR density in general field peaks at $z\sim2$ 
(e.g., \citealt{2013MNRAS.432...23G,2014ARA&A..52..415M,2017MNRAS.467.1360B}),
the SFRs of the protoclusters evaluated by only multiplying the total IR luminosity 
by the conversion factor in \citet{1998ARA&A..36..189K} 
are likely on one track which rapidly evolves at $z=0-0.5$ and 
continues to increase up to $z\sim4$. 
However, if we subtract the emission from AGNs, 
the SFR of a protocluster drops at $z=3\sim4$. 
The protoclusters in the literature also need the consideration of AGNs. 
Though not as much as AGNs, 
the SFR to FIR luminosity relation depends on the assumed stellar population synthesis model. 
It can be said that the total IR luminosity of massive protoclusters continues 
to increase up to $z\sim4$;
however, to show the evolution of the total SFR/SFR density, 
a more careful treatment of AGNs and stellar population of galaxies is needed. 

\subsection{QSOs and protoclusters}

The correlation between the QSOs and protoclusters remains an open issue. 
QSOs are frequently used as land-marks of protoclusters 
however they are not always in overdense regions 
\citep{2009ApJ...695..809K,2017ApJ...841..128K, 2018PASJ...70S..32U,2017MNRAS.470L.117G}.
It has been argued that protoclusters are preferentially found around radio-loud AGNs 
\citep{2014MNRAS.445..280H} while radio-quiet AGNs do not often trace the density peaks, 
except for QSO pairs and multiplets (e.g., \citealt{2018PASJ...70S..31O,2018PASJ...70S..32U}).
Previously, \citet{2018PASJ...70S..32U} showed that only two out 
of the 151 QSOs at $z\sim3.8$ selected from the SDSS survey
are in the protoclusters at $z\sim3.8$ studied here. 

\begin{figure}
\centering
\includegraphics[width=85mm]{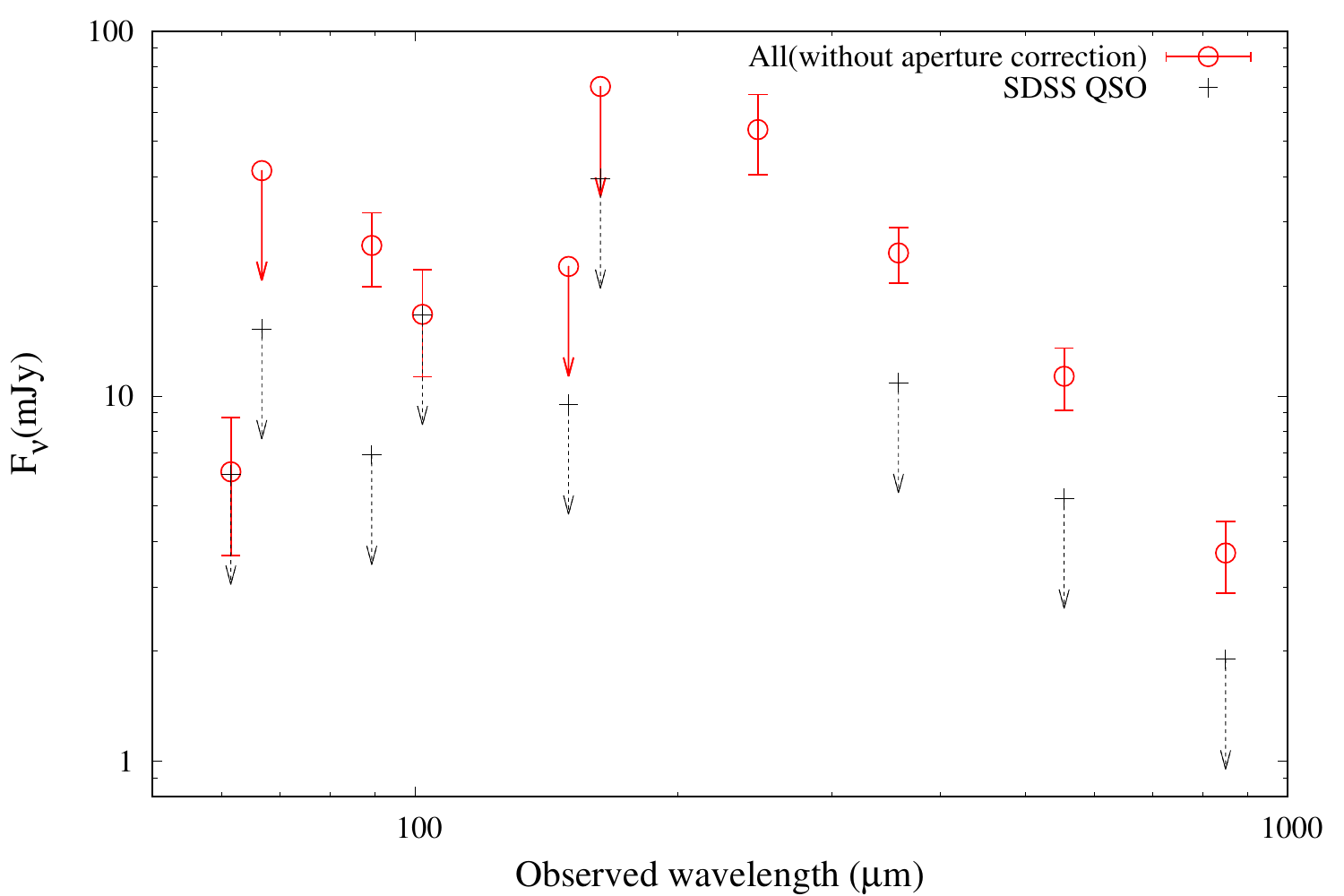}
\caption{The 4-arcmin diameter aperture fluxes measured on $IRAS$, $AKARI$ and $Planck$ stacks 
of the HSC-SSP protoclusters and SDSS QSOs at $z\sim4$. 
The red open circles show the average flux of a protocluster (all).  
The black crosses show the 2$\sigma$ upper limit fluxes of a SDSS QSO.  
}\label{fig:wqso}\end{figure}

We compare the 4-arcmin diameter aperture fluxes measured on $IRAS$, $AKARI$ and $Planck$ stacks 
of all the protoclusters and SDSS QSOs at $z\sim3.8$ in Fig. \ref{fig:wqso}, 
which are measured in the same manner. 
However, SDSS QSOs are not detected at all, although the stacked numbers 
of them are not appreciably different from the protoclusters. 
This supports the results in \citet{2018PASJ...70S..32U} 
that SDSS QSOs at $z\sim3.8$ are not in special regions in general. 
It is also consistent with \citet{2015MNRAS.446.2696S} referred in Section 5.2. 

However, the average MIR luminosity, 
which is an excellent measure of AGN activity, 
of SDSS QSOs is not significantly different
from that of the HzRGs of the known protoclusters; 
The IR SED of Spiderweb HzRG at $z=2.16$ is similar to that of the SDSS QSOs (Fig. \ref{fig:wlbg2}). 
The HzRGs of the other known protoclusters at $z=2-3$ have $X$-ray luminosities 
$L_{2-10 keV}=1\sim4\times 10^{45}$ erg $s^{-1}$ 
\citep{2002A&A...396..109P,2003ApJ...596..105S, 2007MNRAS.376..151J,2019ApJ...874...54M}, 
which corresponds to $\nu L_{\nu}=10^{46}\sim10^{47.5}$ erg $s^{-1}$ 
at $6~\mu$m  ($\approx 0.03\sim1$ mJy at $z=3.8$) 
according to the empirical $L_{2-10 keV}$ v.s. $\nu L_{\nu}$ 
at a 6 $\mu$m relation (e.g., \citealt{2015ApJ...807..129S}).

Meanwhile, the protoclusters at $z\sim3.8$ show strong excess at $<100~\mu$m in the rest-frame
which implies the overdensities and/or enhanced activities 
of obscured AGNs in the protoclusters correspond to $\sim$ ten times that of a SDSS QSO at $z\sim4$ in the IR. 
Interestingly, warm/hot dust emission of the protoclusters becomes luminous approaching $z\sim4$
while the change at $>100~\mu$m is small. 
Assuming that the excess warm/hot dust emission originates in AGNs, 
this implies that the growth of SMBHs in protoclusters peaks at $z\sim4$ or more 
in advance of that in the general field at $z\sim2$ \citep{2014ARA&A..52..415M}. 
If this scenario is true, SDSS QSOs are not good landmarks of protoclusters  at $z\sim4$
because the proto-BCG like sources  at $z\sim4$ are more luminous than them
but buried in dust formed by accompanying intense star formation. 
 
\section{Conclusion}

By stacking {\it Planck}, {\it AKARI}, {\it IRAS}, {\it WISE}, and $H$-ATLAS
images of the largest catalog 
of the protoclusters at $z\sim4$ obtained by the HSC-SSP survey, 
we successfully show their average IR SED for the first time.
The protoclusters at $z\sim4$ are several tens of times brighter than a typical SFG at $z\sim4$. 
They are on average as luminous as the most prominent protoclusters at $z=2-3$ 
and contain a larger warm/hot dust component. 
This suggests that protoclusters have rapidly evolved from $z=2$ to 4. 
The average  IR SED of the protoclusters is unlike the average SED of a typical SFG 
but similar to IR luminous DOGs whose IR emission is dominated by AGNs. 
We evaluate the average SFR of the protoclusters 
by fitting the observed SEDs with SFG and AGN/SFG composite SED models.
For the pure star-forming model, we find $L_{FIR}=19.3_{-4.2}^{+0.6}\times10^{13}~L_{\odot}$ and 
SFR $=16.3_{-7.8}^{+1.0}\times10^3~M_{\odot}$ yr$^{-1}$
while for the AGN/SFG composite model, we find 
$L_{FIR}=5.1_{-2.5}^{+2.5}\times10^{13}~L_{\odot}$ and SFR$=2.1^{+6.3}_{-1.7}\times10^3~M_{\odot}$ yr$^{-1}$.
Their degeneracy cannot be solved via the SED fitting at this point;
however, the contribution from AGNs is not empirically negligible.  
Our results are nearly consistent with previous CIB anisotropy studies,   
but at a shorter wavelength, CIB can be more biased in protocluster regions.
Large uncertainty remained in the total SFR estimates; however,
the total IR luminosity of the most massive clusters are likely to continue increasing up to $z\sim4$. 
Stacking analysis of the QSOs at $z\sim4$ optically selected is also performed 
and no excess star formation around them, as reported in \citet{2018PASJ...70S..32U}, is confirmed. 

Finally, we compare our results to the cosmological simulations of dusty SFGs to date
(e.g., \citealt{2009ApJ...691..560C,2011MNRAS.417.2057A,2013MNRAS.428.2529H,2015MNRAS.450.1320G,2015MNRAS.452..878M, 2016MNRAS.461.1621C,2016MNRAS.462.3854L}).
Simulations predict that SMGs (e.g., with the flux over a few mJy in 850 $\mu$m in their definition) 
are in general strongly biased population hosted in massive halos
with $M_h\gsim10^{11.5}~M_{\odot}$
\citep{2009ApJ...691..560C,2011MNRAS.417.2057A,2013MNRAS.428.2529H,2015MNRAS.452..878M, 2016MNRAS.461.1621C}. 
\citet{2009ApJ...691..560C} and \citet{2015MNRAS.452..878M}
predicted that the density excesses of SMGs do not always trace the most massive protoclusters. 
This agrees with our results that above 4$\sigma$ significance, 
the overdensity significance of $g$-dropout galaxies, 
which more tightly correlates with a halo mass, 
and the total IR luminosity do not correlate well. 
In addition, one or a few IR luminous DOGs such as in \citet{2018ApJ...854..157F} and \citet{2018ApJ...857...31T}
can be responsible for the IR flux of a protocluster.  
If so, a protocluster may not be observed as an significant overdensity of SMGs. 
Simulations also predicted that the peak of the star formation history 
of cluster-sized halos is earlier than that in the general field 
(e.g., \citealt{2017ApJ...844L..23C,2018MNRAS.473.2335M}). 
\citet{2013ApJ...770...57B,2019MNRAS.488.3143B} linked
the galaxy-halo assembly history from simulations 
and the observed galaxy properties, and found that halos with $M_h=10^{14}~M_{\odot}$ at present 
have a star formation history peak at $z\sim3$. 
Our results suggests that the peak of the star formation history can be at $z>4$, 
earlier than that of the predictions using simulations and semi-observational methods. 

Our results demonstrate the great importance of the IR properties of SFGs 
and AGNs in protoclusters ``typical" at $z\sim4$, for the first time. 
On the whole, our results suggest that DSFGs in protoclusters at $z\sim4$ 
are more common than those predicted by current simulations. 
According to our results, simulations will need to approach 
the statistical behavior of the richest clusters with $M_h>10^{14}~M_{\odot}$ with a larger simulation box, 
dust emissivity at mid to far-IR, 
role of AGNs, and further constraints on the evolution of protoclusters at $z>3$ in the future. 
From the observational side, 
we can expand our study with HSC-SSP and next generation telescopes. 
Notably, the catalog used here is only a tenth of the whole HSC-SSP WIDE layer.
In addition, Large Synoptic Survey Telescope 
will provide an additional large catalog of protoclusters in the future. 
These surveys will enable deeper stacking analysis 
for protoclusters at various redshifts.  
Characterization of individual IR sources
in protoclusters is also needed 
though it is beyond the scope of this paper.
\citet{2019ApJ...878...68I} optically selected the predominantly 
bright sources in some of our protocluster candidates
as the candidate brightest cluster galaxies (BCGs). 
They are among the possible sources dominating the IR emission. 
Our study provides an excellent simulation for the James Webb Space Telescope ({\it JWST}) 
and Space Infrared Telescope for Cosmology and Astrophysics ({\it SPICA}). 
At this point, the deep observations with $8-10$ m class telescopes in the NIR, 
ALMA, $Chandra$, and $XMM-Newton$ telescopes
are feasible to identify DSFGs/AGNs of protoclusters. 
However, given their large variation, several protoclusters at $z\sim4$ need to be observed to evaluate a typical value. 

%----------------------
\acknowledgments
NK acknowledges support from the JSPS grant 15H03645.
RAO is grateful for financial support from the S\~ao Paulo Research Foundation (FAPESP; grant 2018/02444-7). 
YM acknowledges support from the JSPS grants 17H04831, 17KK0098 and 19H00697.
The HSC collaboration includes the astronomical communi- ties of Japan and Taiwan, and Princeton University. The HSC instrumentation and software were developed by the National Astronomical Observatory of Japan (NAOJ), the Kavli Institute for the Physics and Mathematics of the Universe (Kavli IPMU), the University of Tokyo, the High Energy Accelerator Research Organization (KEK), the Academia Sinica Institute for Astronomy and Astrophysics in Taiwan (ASIAA), and Princeton University. Funding was contributed by the FIRST program from the Japanese Cabinet Office, the Ministry of Education, Culture, Sports, Science and Technology (MEXT), the Japan Society for the Promotion of Science (JSPS), the Japan Science and Technology Agency (JST), the Toray Science Foundation, NAOJ, Kavli IPMU, KEK, ASIAA, and Princeton University. This paper makes use of software developed for the Large Synoptic Survey Telescope (LSST). We thank the LSST Project for making their code available as free software at http://dm.lsst.org.
This research is based on observations with {\it AKARI}, a JAXA project with the participation of ESA.
This research uses {\it WISE} Release data is eligible for proposals to the NASA ROSES Astrophysics Data Analysis Program.
This research has made use of the NASA/ IPAC Infrared Science Archive, which is operated by the Jet Propulsion Laboratory, California Institute of Technology, under contract with the National Aeronautics and Space Administration.
And many thanks to the oversea program of National Astronomical Observatory of Japan.

\newpage
\appendix
\section{DATA SUMMARY}
\restartappendixnumbering
\begin{deluxetable*}{cccccc}[b!]
%\tablenum{1}
\tablecaption{Summary of the data\label{tab:data}}
\tablewidth{0pt}
\tablehead{
\colhead{Instrument} & \colhead{Band} & \colhead{W$_{cen}$$^a$} & \colhead{FWHM PSF$^a$} & \colhead{Point source detection limit$^b$} & \colhead{1$\sigma$ (stack, 4')$^c$} \\
 \colhead{}   &  \colhead{} &  \colhead{($\mu$m)}   &  \colhead{(arcmin)}   &  \colhead{(Jy)	}   &  \colhead{(mJy)}	 
}
\decimalcolnumbers
\startdata
{\it Planck}$^a$ & 857 GHz & 350	&	4.92	&	0.166	&	5.2	\\	
 & 545 GHz & 540	&	4.68	&	0.118	&	2.8	 \\	
 & 353 GHz & 840	&	4.22	&	0.069	&	1.1	 \\
{\it IRAS}    &  60 &	 60	  &	3.6   &	0.6	   &	4.7	\\	
	     & 100 &	100	  &	4.2   &	1.0	   &	7.0	\\
\enddata
\tablecomments{$^a$The central wavelengths and FWHM of the PSFs for $Planck$ are from https://wiki.cosmos.esa.int/planck-legacy-archive/index.php. For $IRAS$, $AKARI$, $WISE$ and $Herschel$, we refer $IRAS$ Explanatory Supplement (https://irsa.ipac.caltech.edu/IRASdocs/exp.sup),  \citet{2015PASJ...67...50D} and \citet{2015PASJ...67...51T},  http://wise2.ipac.caltech.edu/docs/release/allsky and https://www.cosmos.esa.int/web/herschel/home, respectively. For WISE, we list FWHM of the PSFs for Atlas image which are larger than those of a single exposure image.
$^b$The point source detection limits in the literatures. For {\it Planck}, we refer the 90 \% completeness limits listed in Table 1 of \citet{2016AA...596A.100P}, originally given in mJy. For {\it IRAS}, we refer the completeness limit for {\it IRAS} Point Source Catalog, Version 2.0 (\citealt{1988iras....7.....H}; https://heasarc.gsfc.nasa.gov/W3Browse/iras /iraspsc.html). For {\it AKARI}, we refer {\it AKARI}/FIS All-Sky Survey Bright Source Catalogue Version 1.0 Release Note (https://www.ir.isas.jaxa.jp/AKARI/Archive/Catalogues/PSC/RN/AKARI-FIS\_BSC\_V1\_RN.pdf). For {\it Herschel}, we refer the $5 \sigma$ detection limit for {\it H}-ATLAS data release 1 \citep{2016MNRAS.462.3146V}, originally given in mJy. For {\it WISE}, we refer the $5 \sigma$ detection limit in the Release Note (http://wise2.ipac.caltech.edu/docs/release/allsky).
$^c$The standard deviation of the fluxes in 4-arcmin diameters measured by 1000 times iteration of the stacking analysis at random positions in similar manner with the protoclusters ($N=216$ for {\it Planck}, {\it IRAS}, {\it AKARI} and  {\it WISE}, and $N=93$ for $H$-ATLAS). }
\end{deluxetable*}
\begin{deluxetable*}{cccccc}
\tablenum{A1}
\tablecaption{Summary of the data --continues}
\tablewidth{0pt}
\tablehead{
\colhead{Instrument} & \colhead{Band} & \colhead{W$_{cen}$$^a$} & \colhead{FWHM PSF$^a$} & \colhead{Point source detection limit$^b$} & \colhead{1$\sigma$ (stack, 4')$^c$} \\
 \colhead{}   &  \colhead{} &  \colhead{($\mu$m)}   &  \colhead{(arcsec)}   &  \colhead{(Jy)	}   &  \colhead{(mJy)}	 
}
\decimalcolnumbers
\startdata
{\it AKARI}    &    {\it N60}	&	65	  &	63.4   &	2.4	   &	16	 \\	
		&    {\it WIDE-S}	&	90	  &	77.8   &	0.5	   &	5.1		 \\	
		&    {\it WIDE-L}	&	140	  &	88.3   &	1.4	   &	7.1		 \\	
		&    {\it  N160}	&	160	  &	88.3$^d$   &	6.3	   &	14	 \\	
{\it Herschel}     &    100	&	98	 	 &	$11.8\times11.0$  &	0.220	   &	64	 \\	
		    &    160	&	154	  &	$14.6\times12.9$   &	0.245	   &	48	 \\	
		    &    250	&	247	  &	$18.4\times17.4$   &	0.037	   &	20	 \\	
		    &    350	&	347	  &	$24.9\times23.6$   &	0.047	   &	6.5	 \\	
		    &    500	&	497	  &	$37.0\times33.8$   &	0.051   &	4.8	 \\	
{\it WISE}     &    W1 & 3.368 &	8.25  &	0.068 $\times10^{-3}$   &	0.56	 \\	
		    &    W2	&	4.618	  &	8.25   &	0.098 $\times10^{-3}$	   &	0.27	 \\	
		    &    W3	&	12.082	  &	8.25   &	0.86 $\times10^{-3}$	   &	0.16	 \\	
		    &    W4	&	22.194	  &	16.5  &	5.4 $\times10^{-3}$	   &	0.27	 \\	
\enddata
\tablecomments{$^d$FWHM of the PSF for {\it WIDE-L}}
\end{deluxetable*}

\section{Stacking analysis simulation}

We simulate the flux of a protocluster enclosed in an aperture by stacking mock images. 
Here four cases are assumed as follows: 
(1) one source at a random position within 2 arcmin, 
(2) three to five sources at random positions and flux values within 2.5 arcmin, 
(3) three to five sources at random positions and flux values within 5.0 arcmin from the center, and 
(4) observed distribution and fluxes of SMGs at $z_{\rm spec}\approx3.09$ in the SSA22 protocluster.
Their average flux distributions are more extended in the order of (1)$<$(2)$<$(4)$<$(3). 
At total of  214 mock images are generated and, smooth the images 
to have FWHM PSF 4.9 arcmin;, and they are stacked for each case.
The total flux of the sources on a mock image is a fixed value.
In case (4), random rotations and random shifts within 2 arcmin 
centering at the brightest source are added. 

Fig. \ref{fig:sim1} shows the simulated images. 
We compare the observed radial profiles of the protoclusters 
at 857 GHz smoothed to have a FWHM of the PSF 4.9 arcmin
to the simulations in Fig. \ref{fig:sim2} ({\it left}).
The observed radial profiles are more extended than that of a point source. 
All protoclusters are similar to case (1) and (2) while 
the brighter-half protoclusters are more similar to case (4).  
Fig. \ref{fig:sim2} ({\it right}) shows the flux fraction enclosed in an aperture. 
In cases (1), (2) and (4), $\sim29\%,~28\%$, and $22$~\% of the total fluxes 
are enclosed in a 2 arcmin aperture radius. 
The aperture correction factors obtained using {\it Herschel} as in Section 3.4 are identical to them.

\setcounter{figure}{0}
\begin{figure*}\centering
\includegraphics[width=44mm]{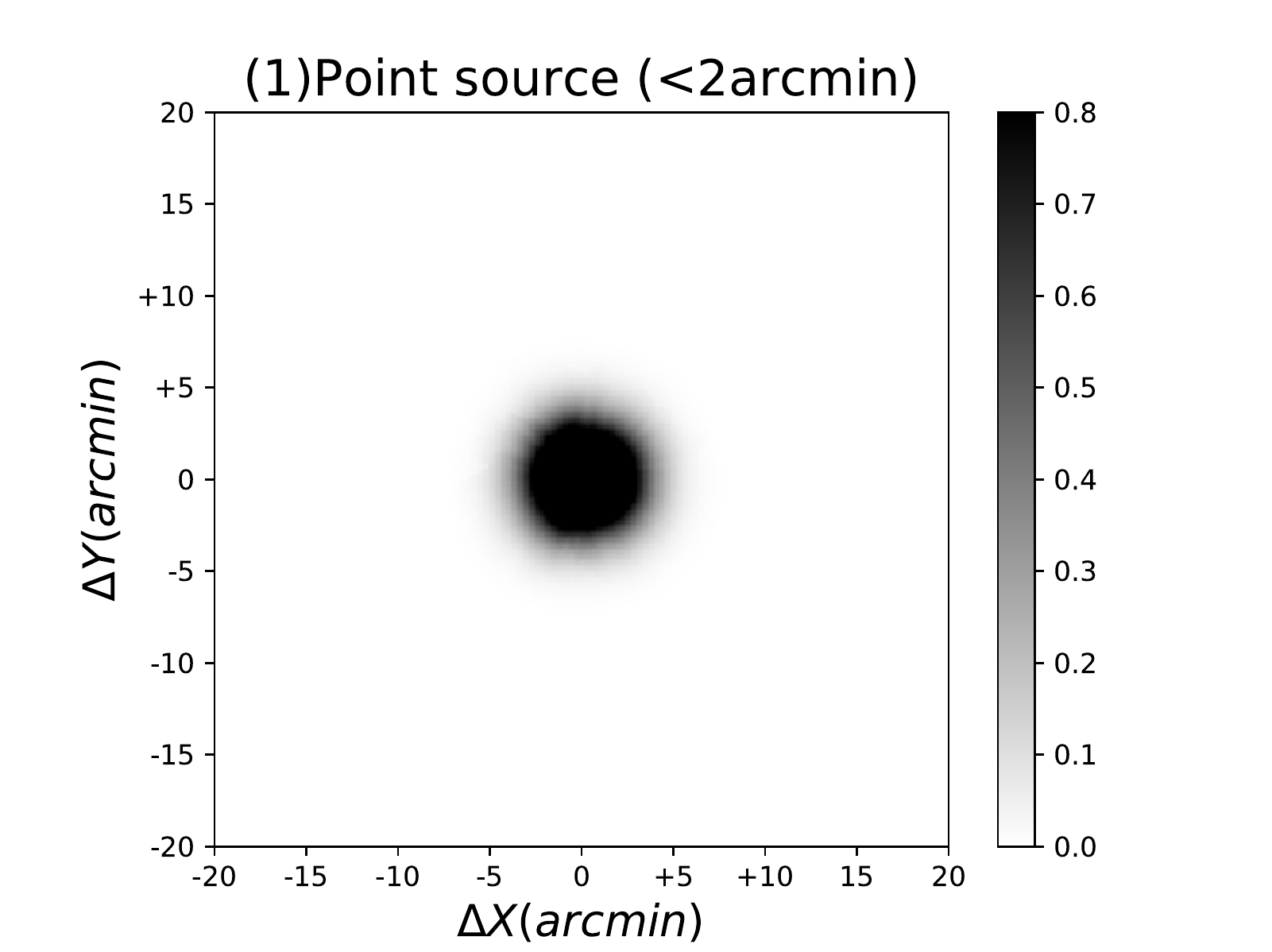}
\includegraphics[width=44mm]{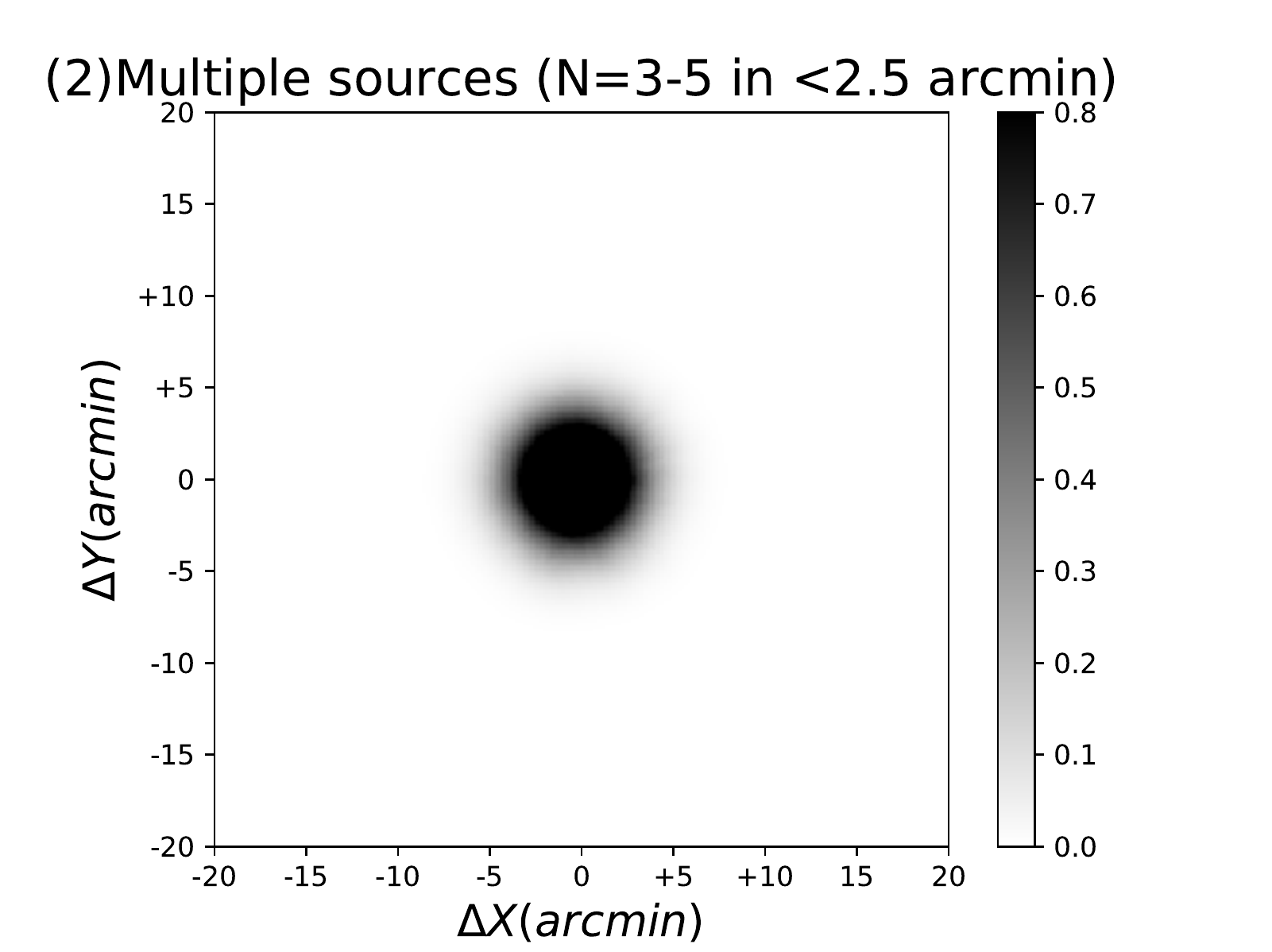}
\includegraphics[width=44mm]{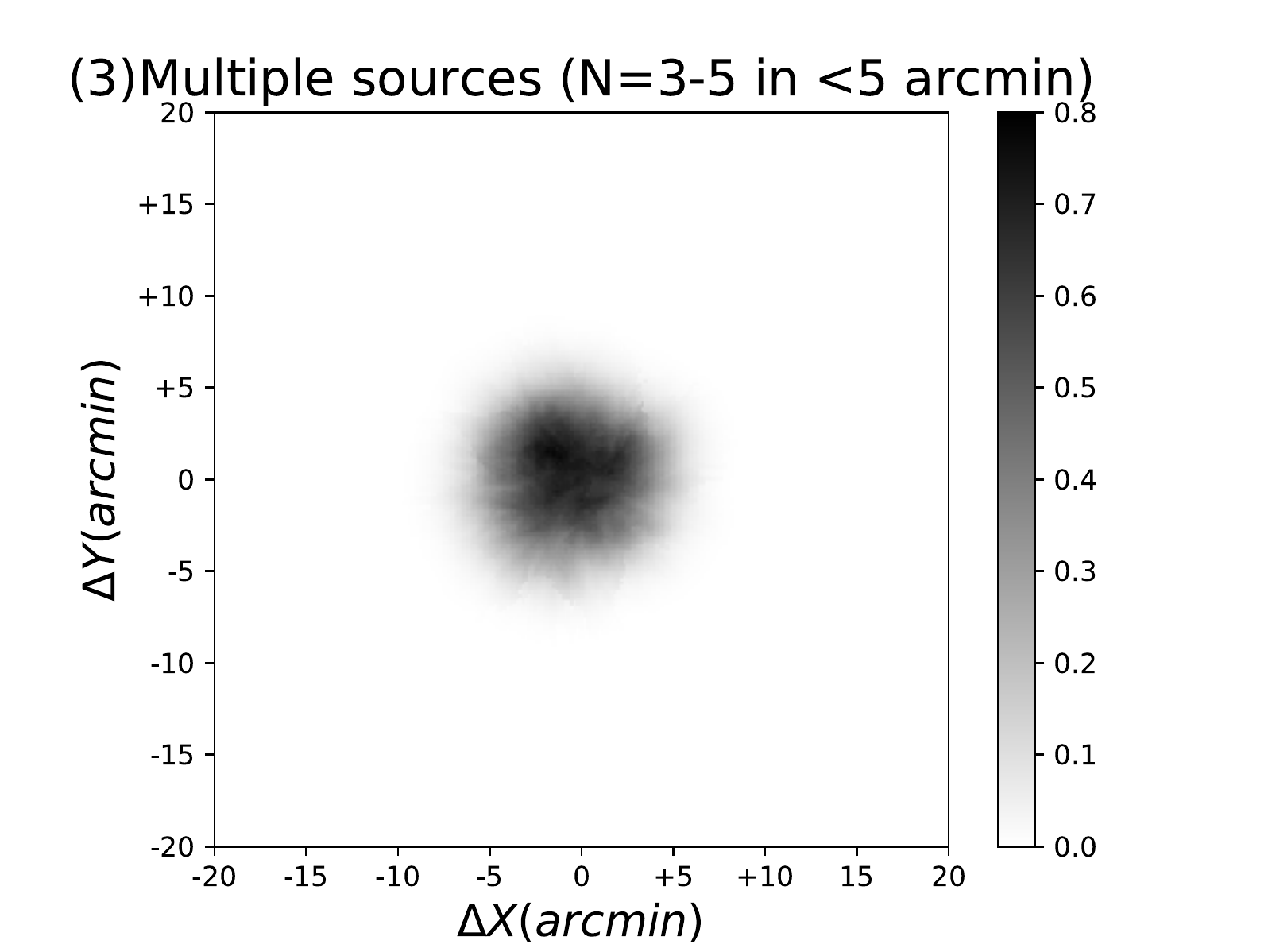}
\includegraphics[width=44mm]{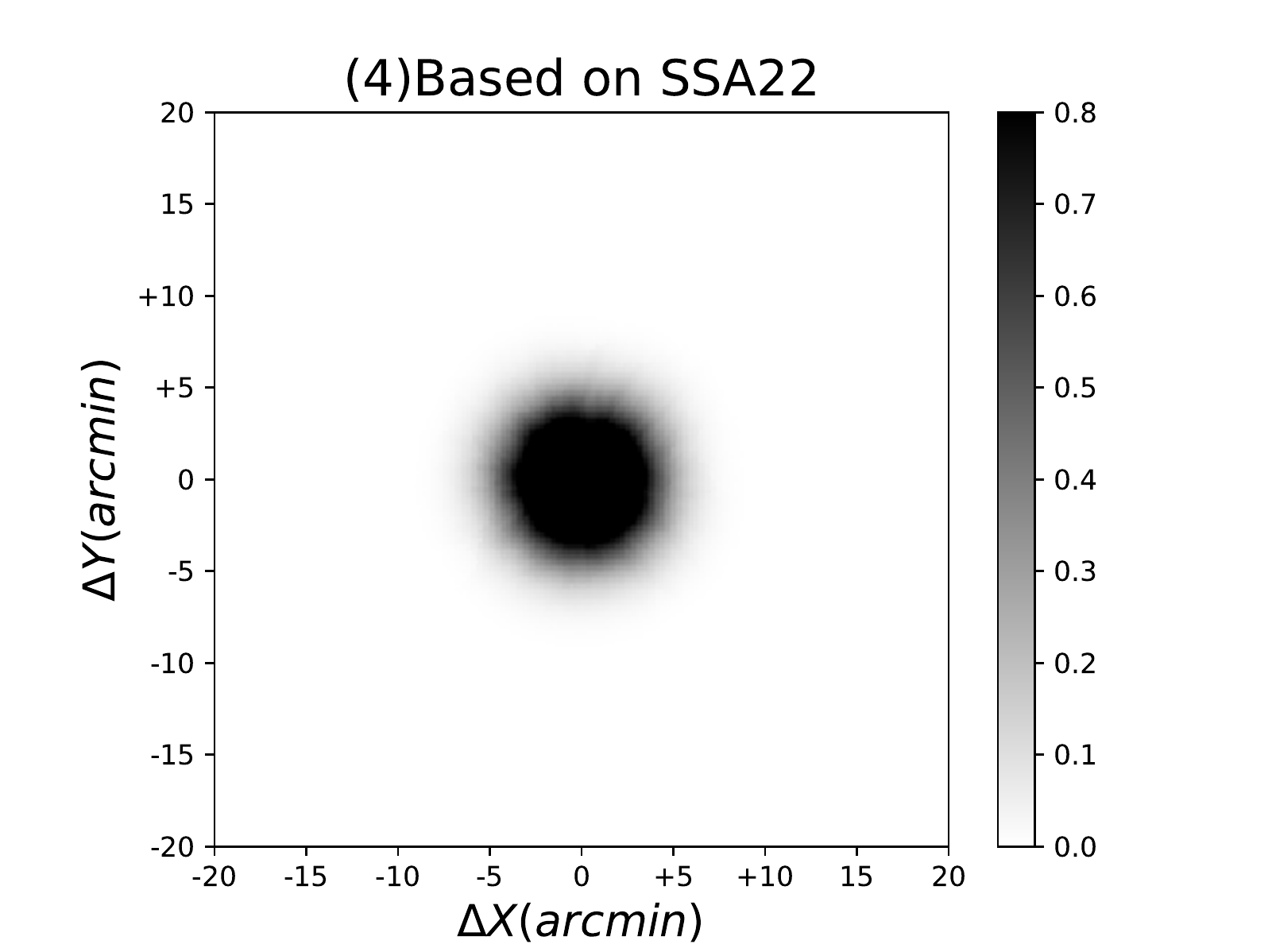}
\caption{ Simulation of the stacking analysis in case (1) to (4) from {\it left} to {\it right}}
\label{fig:sim1} \end{figure*}

\begin{figure*}\centering
\includegraphics[width=80mm]{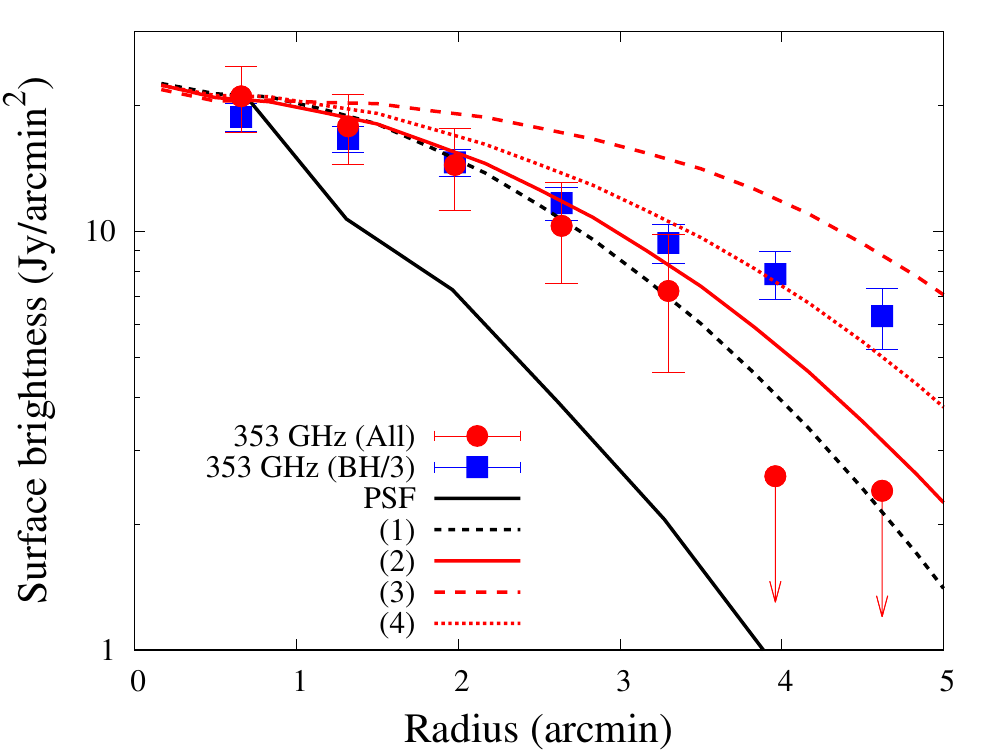}
\includegraphics[width=80mm]{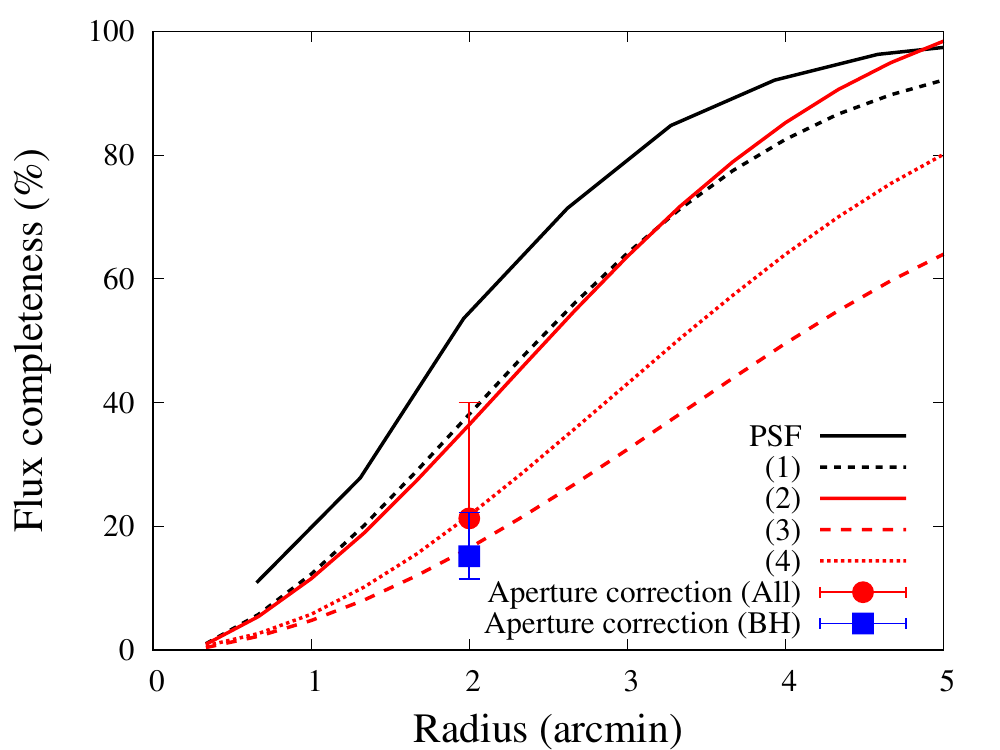}
\caption{ {\it Left:} Radial profiles of the protoclusters compared with the simulations. 
The black solid curve shows the radial profile of a point source. 
The black dashed curve, and red solid, dashed, and dotted curves
show the average radial profiles for case (1), (2), (3), and (4) 
measured by a thousand times iterations for each. 
The red filled circles and blue filled squares show the radial profiles 
of the all and brighter-half protoclusters measured on the stacked 
images at {\it Planck} 857 GHz smoothed to have FWHM of the PSF 4.9 arcmin. 
{\it Right:} Similar to {\it left} panel but flux completenesses at a given aperture are shown in $y$-axis.  
The red filled circle and blue filled square show the aperture correction factors found in Section 3.4. }
\label{fig:sim2} \end{figure*}

\section{Fainter half protoclusters}
Figures \ref{fig:fh1} and \ref{fig:fh2} 
show the $Planck$ 353, 545, and 857-GHz stacks for the 1st quartile and 2nd quertile 
from the lowest of the 857 GHz flux distribution of the protoclusters  (Fig. \ref{fig:fluxdist}).

\setcounter{figure}{0}
\begin{figure*}\centering
\includegraphics[width=54mm]{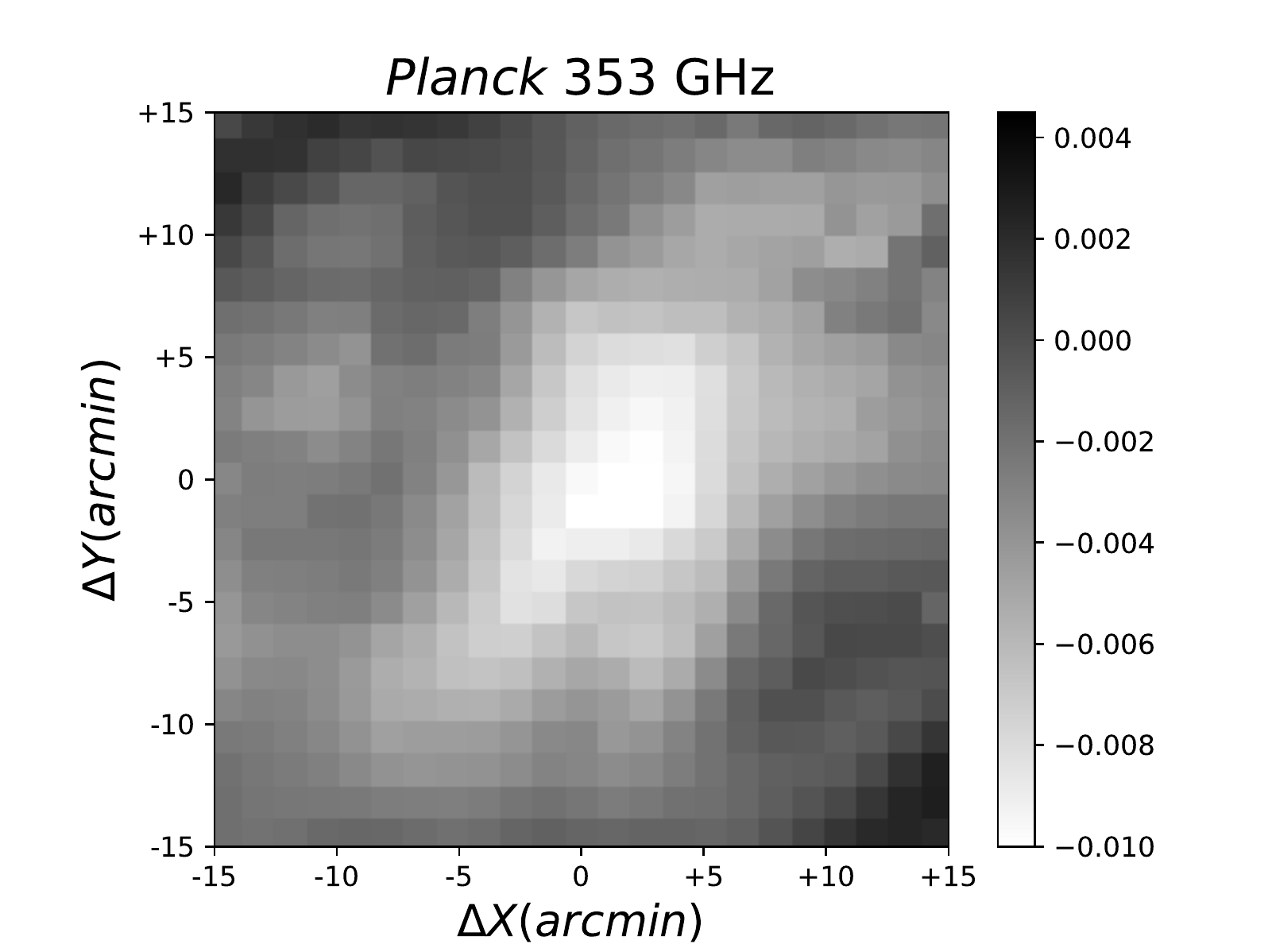}	\hspace{-18pt}
\includegraphics[width=54mm]{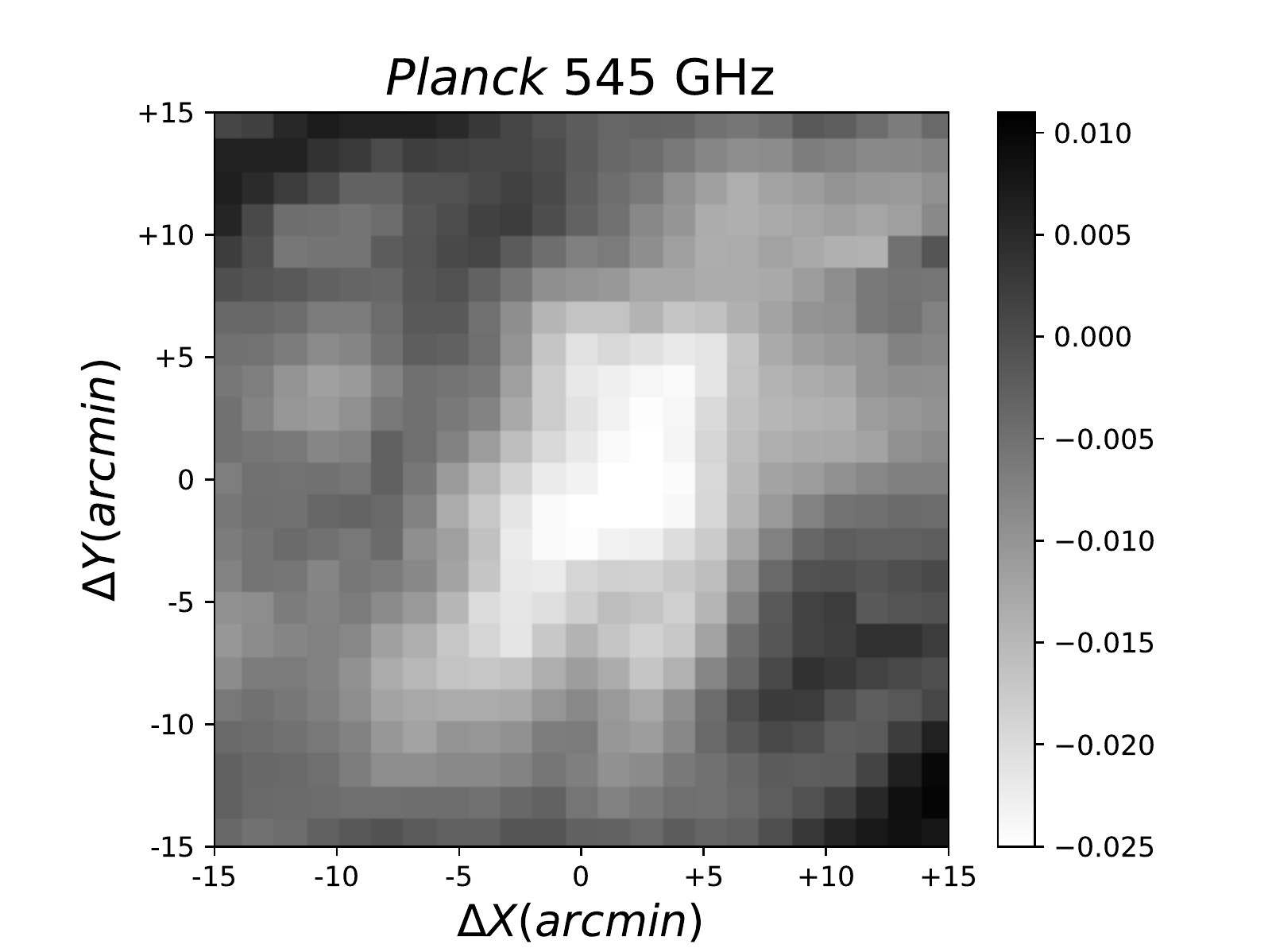}	\hspace{-18pt}
\includegraphics[width=54mm]{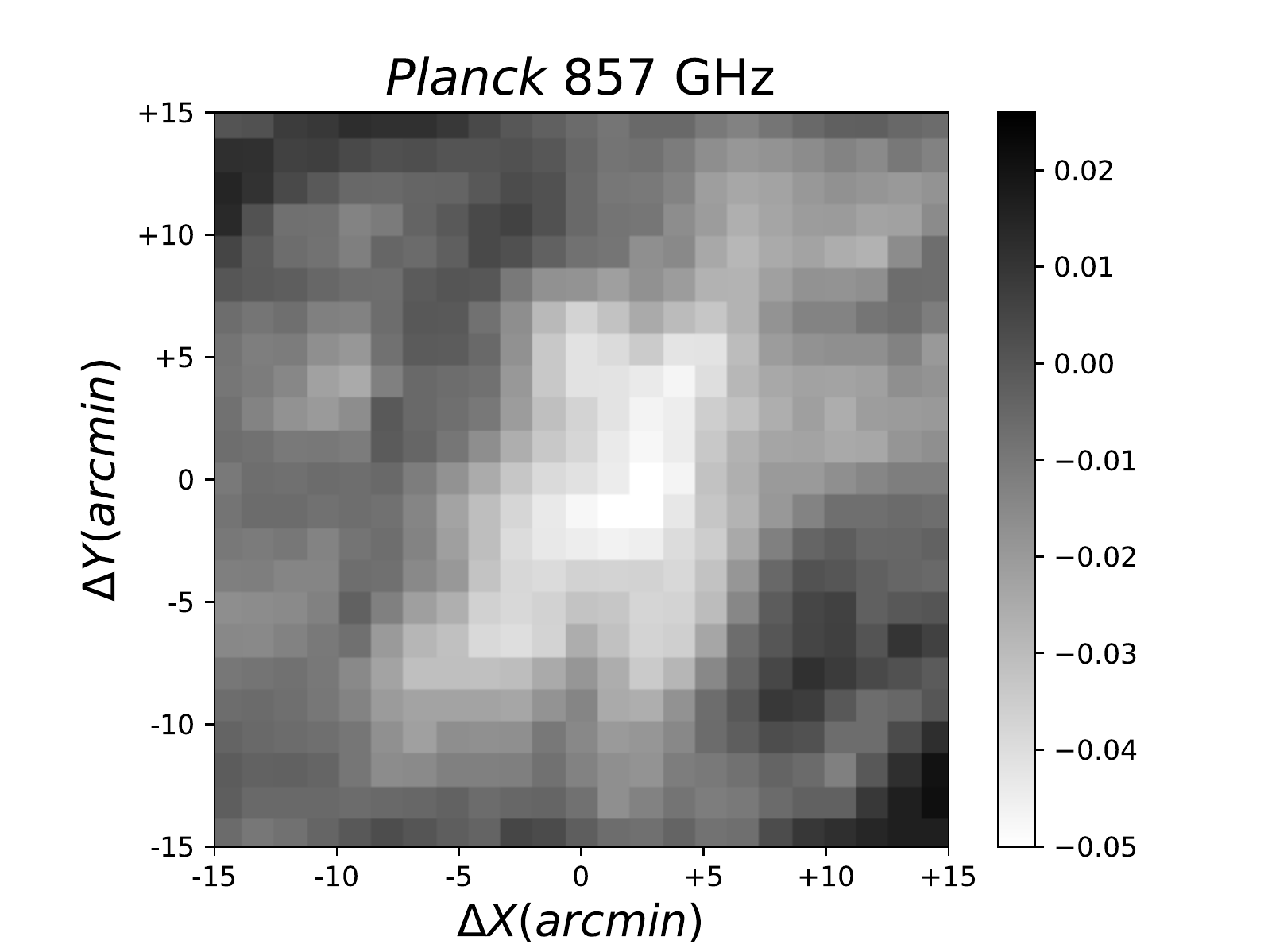}	\hspace{-18pt}
\caption{ $Planck$ 353, 545, and 857-GHz stacks for the 1st quartile from the lowest of the flux distribution of the protoclusters, from left to right. }
\label{fig:fh1}
\includegraphics[width=54mm]{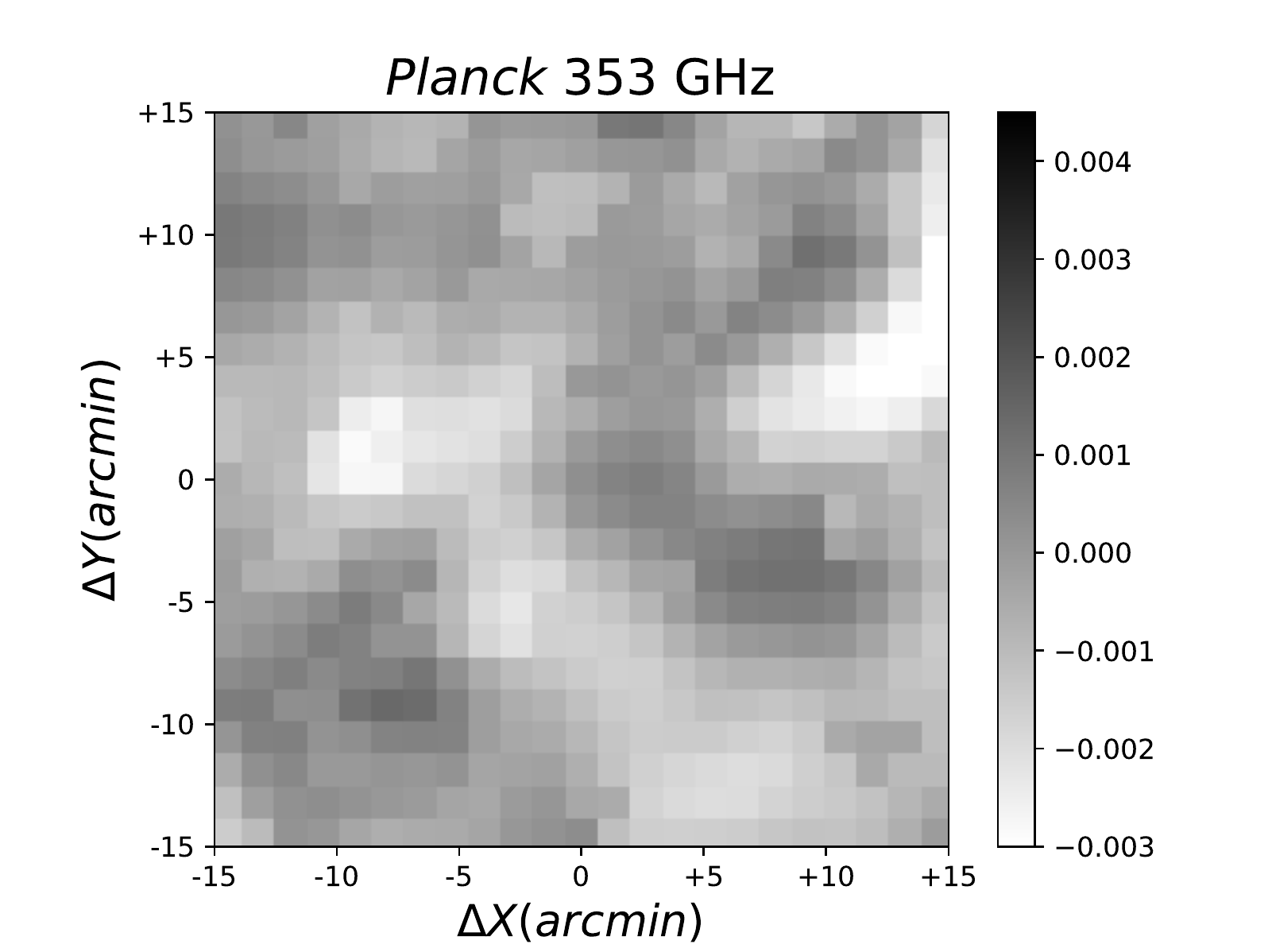}	\hspace{-18pt}
\includegraphics[width=54mm]{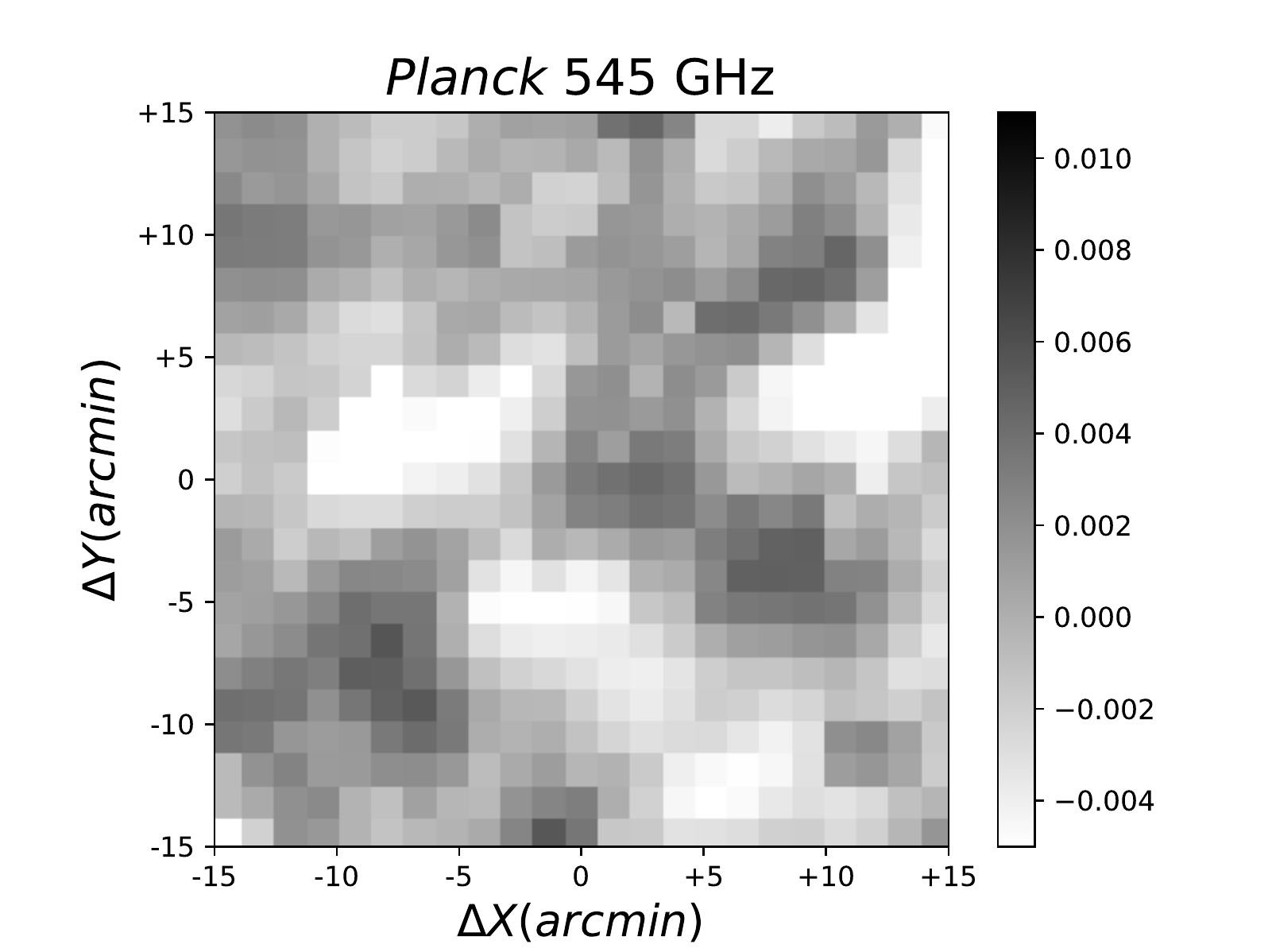}	\hspace{-18pt}
\includegraphics[width=54mm]{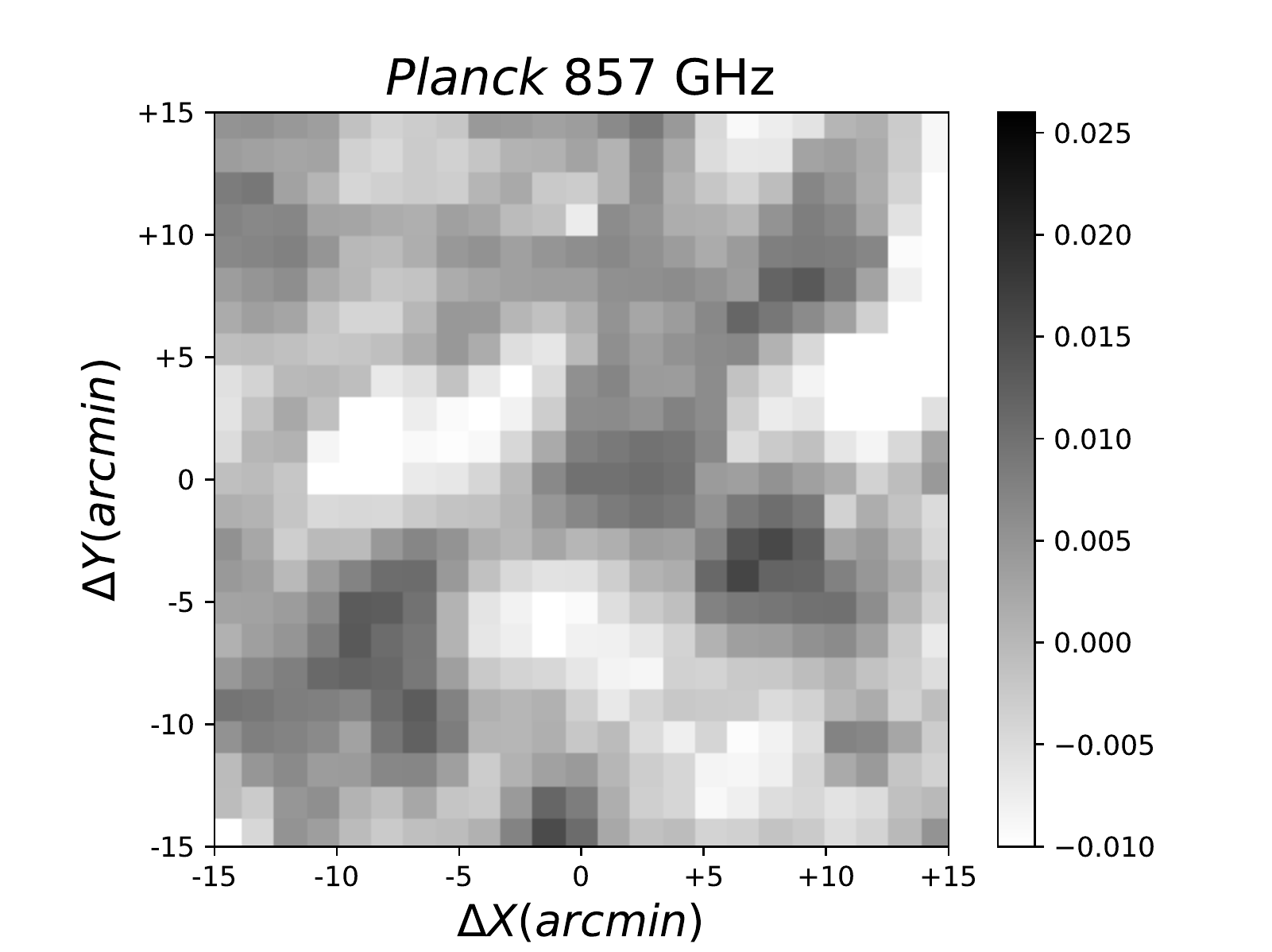}	\hspace{-18pt}
\caption{ $Planck$ 353, 545, and 857-GHz stacks for the 2nd quartile from the lowest of the flux distribution of the protoclusters, from left to right. }
\label{fig:fh2} \end{figure*}

\end{document}